\documentclass[a4paper,11pt]{article}
\pdfoutput=1
\usepackage{jcappub}

\usepackage{pdflscape}
\usepackage{graphicx}
\usepackage{multirow}
\usepackage{amsmath}
\usepackage{float}
\usepackage[normalem]{ulem}

\usepackage{xcolor}
\usepackage{csquotes}
\usepackage{array}

\usepackage[range-phrase=\textendash, range-units=single]{siunitx}
\let\sun\odot
\DeclareSIUnit\solarmass{\ensuremath{M_\sun}}
\DeclareSIUnit\year{yr}
\DeclareSIUnit\erg{erg}
\DeclareSIUnit\parsec{pc}
\DeclareSIUnit\c{c}

\graphicspath{{figures/}}

\title{Searches for coincident Gravitational Waves and High-Energy Neutrinos from the KM3NeT and ANTARES telescopes during O3}

\collaboration{The ANTARES and KM3NeT Collaborations}
\author[a,b]{O. Adriani}
\author[c,d]{A. Albert}
\author[e]{A. R. Alhebsi}
\author[e]{S. Alshalloudi}
\author[f]{S. Alves}
\author[g]{F. Ameli}
\author[f]{F. Andersen}
\author[h]{M. Andre}
\author[i]{L. Aphecetche}
\author[j]{M. Ardid}
% ...

\emailAdd{km3net-pc@km3net.de}
\emailAdd{duverne@apc.in2p3.fr}
\emailAdd{lamoureux@apc.in2p3.fr}
\emailAdd{pillas@iap.fr}
\emailAdd{iara.tosta.melo@dfa.unict.it}

\affiliation[a]{INFN, Sezione di Firenze, via Sansone 1, Sesto Fiorentino, 50019 Italy} 
\affiliation[b]{Università di Firenze, Dipartimento di Fisica e Astronomia, via Sansone 1, Sesto Fiorentino, 50019 Italy} 
\affiliation[c]{Université de Strasbourg, CNRS, IPHC UMR 7178, F-67000 Strasbourg, France} 
\affiliation[d]{Université de Haute Alsace, rue des Frères Lumière, 68093 Mulhouse Cedex, France} 
\affiliation[e]{Khalifa University of Science and Technology, Department of Physics, PO Box 127788, Abu Dhabi, United Arab Emirates} 
\affiliation[f]{IFIC - Instituto de Física Corpuscular (CSIC - Universitat de València), c/Catedrático José Beltrán, 2, 46980 Paterna, Valencia, Spain} 
\affiliation[g]{INFN, Sezione di Roma, Piazzale Aldo Moro, 2 - c/o Dipartimento di Fisica, Edificio, G.Marconi, Roma, 00185 Italy} 
\affiliation[h]{Universitat Politècnica de Catalunya, Laboratori d'Aplicacions Bioacústiques, Centre Tecnològic de Vilanova i la Geltrú, Avda. Rambla Exposició, s/n, Vilanova i la Geltrú, 08800 Spain} 
\affiliation[i]{Subatech, IMT Atlantique, IN2P3-CNRS, Nantes Université, 4 rue Alfred Kastler - La Chantrerie, Nantes, BP 20722 44307 France} 
\affiliation[j]{Universitat Politècnica de València, Instituto de Investigación para la Gestión Integrada de las Zonas Costeras, C/ Paranimf, 1, Gandia, 46730 Spain} 
% ...

\abstract{Various models predict that gravitational-wave sources produce high-energy neutrinos. Among them are compact binary coalescences with at least one neutron star, but also core-collapse supernovae or accretion disk instabilities. Two different analyses are presented, combining data from the LIGO and Virgo gravitational-wave detectors with those from the ANTARES and ORCA neutrino telescopes observed during the third observing run of LIGO-Virgo. In the first analysis, selected high-energy neutrino candidates were used as triggers to search for gravitational-wave candidates. In the second analysis, a search was conducted for potential spatial and temporal coincidences between subthreshold high-energy neutrino candidate events and subthreshold gravitational-wave candidates emitted by compact binary coalescence systems. In both searches, no significant coincidences were found. Upper limits on the rate of sources emitting both high-energy neutrinos and gravitational waves are presented. In particular, with the joint search, a joint source rate density upper limit of $8.8\times10^2~\mathrm{ Gpc^{-3}yr^{-1}}$ was found for a typical binary neutron star coalescence with a GW emitted energy of $10^{-1}~\rm M_\odot c^2$ and an isotropic energy emitted in neutrinos of $10^{51}~$erg.}

\begin{document}

\maketitle

\section{Introduction}
\label{sec:intro}

Since the first direct detection of gravitational waves (GWs) in 2015 \citep{2016PhRvL.116f1102A}, GW astronomy has become an essential channel for probing compact-object dynamics and for exploring extreme astrophysical environments. Current ground-based interferometers include the two LIGO \citep{LIGOScientific:2014pky} instruments, Hanford and Livingston, Virgo \cite{lvk, VIRGO:2014yos}, and KAGRA \cite{akatsu2020} which have observed hundreds of compact binary coalescences (CBCs) during several data-taking periods \citep{2019PhRvX...9c1040A,2021PhRvX..11b1053A,o3_dist,2025arXiv250818082T}. CBCs are currently the only type of GW sources confidently identified by the LIGO-Virgo-KAGRA (LVK) collaboration. They consist of the merger of either two black holes (BBH), two neutron stars (BNS), or one neutron star and one black hole (NSBH).
A major milestone of GW astronomy was achieved on August 17$^{\rm th}$, 2017, when the GW170817 signal \cite{gw170817}, emitted by the merger of two neutron stars, was detected in coincidence with a gamma-ray burst (GRB) observed by the Fermi satellite \cite{grb17}. This coincidence launched a worldwide follow-up campaign that led to the discovery of electromagnetic (EM) counterparts across the whole spectrum \citep{followup}: the first kilonova in the optical domain \citep{Soares_Santos_2017,at2017gfo}, and the X-ray \citep{Troja_2017} and radio afterglows of the GRB \citep{Hallinan_2017}. The joint detection of GW170817 with these counterparts firmly established the foundations of multi-messenger astronomy with GWs, demonstrating how combining information from different messengers enhances our understanding of astrophysical events. 
      
Although no neutrino counterpart to GW170817 was observed by the ANTARES, IceCube, and Pierre Auger experiments \cite{gwhen17}, there are theoretical models predicting the emission of high-energy neutrinos (HENs) in coincidence with GWs. Among CBC sources, BNS and NSBH mergers are expected to produce HENs through photo-hadronic ($p\gamma$) interactions within the relativistic jet launched after coalescence, yielding charged pions that decay into muons and neutrinos \cite{GRB_jet1, GRB_jet2, GRB_jet3}. For BBH mergers embedded in active galactic nucleus (AGN) accretion disks, neutrino production can arise from hadronic interactions between the post-merger jet and the surrounding disk material \cite{agn_hen_gw}. The potential for neutrino emission from CBCs has motivated a decade of joint GW-HEN analyses. Searches have been performed combining LVK data with neutrino observations from IceCube~\cite{IceCube:2022mma,IceCube:2023atb,abbasi2021probingneutrinoemissiongev}, ANTARES \cite{ANTARES:2013tra,Albert_2023}, and KM3NeT \cite{KM3NeT:2023cdr}. These analyses searched for neutrino emission from well-identified gravitational-wave emitters: BBH, BNS, and NSBH. As no statistically significant excess of neutrinos associated with GW events has been identified so far, upper limits on neutrino fluences and on the fraction of CBCs capable of producing detectable HENs have been reported for the objects investigated. Dedicated analyses focusing on BBHs in AGN environments have also been conducted \cite{Ricca:2025afj,bbh_agn_gw,agn_hen_gw}.
        
Beyond compact object mergers, a wide variety of GW sources are expected to produce signals that are not accurately described by CBC waveform models. These include core-collapse supernovae (CCSNe), magnetar flares, post-merger remnants, or cosmic string cusps, for example. As a result, their detection relies on more generic, model-independent search strategies, commonly referred to as \emph{unmodeled} or \emph{burst} searches, such as \textit{coherent WaveBurst} (cWB) \citep{2021SoftX..1400678D}. These burst searches can also be applied to look for GW-HEN joint detections from a wider range of sources, such as cosmic strings, CCSNe, or microquasars, for which models predict the emission of both messengers. For CCSN events \citep{supernova1,supernova2}, the joint emission of GWs and HENs arises from the dynamic mechanisms of the collapse, bounce, and shocks \citep{vartanyan2023}. In microquasars, GWs are expected to be emitted during the accretion process as the matter from the companion star falls toward the compact object, converting the gravitational potential into kinetic energy and powering the relativistic jet that characterizes the system \citep{Done_2007}. The jet has a hadronic component \citep{pradier} accompanied by neutrino emission produced by the decay of pion mesons from $pp$ or $p\gamma$ interactions \citep{pradier,Bosch_Ramon_2005}.
Cosmic strings formed in the early universe during symmetry-breaking phase transitions can also be sources of multiple messengers. Due to their kinks and cusps, cosmic strings can generate burst-like GWs, and similarly, they can produce particles, including neutrinos, making them an interesting source in the multi-messenger landscape \citep{cuspHEN,cupsGW}.
        
In this paper, we present the methods and the results of two complementary searches for joint GW--HEN sources with GW and HEN data produced during the third data-taking run of LVK (O3 -- from April 1$^\mathrm{st}$ 2019 to March 27$^\mathrm{th}$ 2020, with a break in October 2019). We make use of the HEN data produced by two neutrino detectors active during O3 in the depths of the Mediterranean Sea: ANTARES and KM3NeT/ORCA. The ANTARES neutrino telescope~\cite{ANTARES:2011hfw}, off the shore of Toulon (France), was operated in its final configuration between May 2008 and February 2022. It was composed of an array of 885 photomultiplier tubes (PMTs) enclosed in pressure-proof glass spheres, arranged over 12 vertical lines and optimized for the detection of the Cherenkov light produced by neutrino interactions in or in the vicinity of the instrument. The KM3NeT infrastructure~\cite{KM3Net:2016zxf} is a second-generation neutrino telescope, currently under construction, and hosts two detectors: ARCA, located off Portopalo di Capo Passero (Italy), optimized for the detection of cosmic neutrinos with energies larger than about 1 TeV, and ORCA, located off Toulon, optimized for the study of atmospheric neutrino oscillations. In their final configurations, the two telescopes will consist of about 300 detection units, each equipped with 18 optical modules, with 31 PMTs each. During O3, KM3NeT recorded physics data in a partial configuration, which included four and then six detection units at the ORCA site. Only the second period is considered and referred to as ORCA6 in the following. In this article, we restrict our work to the ANTARES and ORCA6 data covering the third GW observing run O3. KM3NeT/ARCA was not considered for this analysis, as only one detection unit was taking data during O3. The selected periods correspond to a livetime of 266 days for ANTARES and 45.5 days at the end of the second half of O3 (O3b) for KM3NeT.

The first analysis is a triggered search based on the \texttt{X-pipeline} algorithm \cite{Sutton_2010}, which looks for GWs associated with HENs confidently detected by ANTARES, without assuming a gravitational waveform model (i.e., unmodeled search). \texttt{X-pipeline} is a software package that combines the data from a network of GW interferometers coherently for detecting minimally modeled GW bursts associated with external triggers such as HENs or GRBs. It searches for excess power in the time-frequency domain while imposing consistency conditions across detectors, and it incorporates the HEN trigger information to constrain the sky location and time window of the search. This approach enhances sensitivity to weak signals and provides robust background rejection in the presence of non-Gaussian noise.
        
The second method is a subthreshold search based on the joint search for GW and GRB associations in LIGO/Virgo and Fermi/GBM data~\cite{Ranking,Pillas_2023}, and on a joint GW-HEN neutrino search in ANTARES, IceCube, and Advanced LIGO O1 data~\citep{ANTARES:2018bmu}. Instead of considering high-significance (\emph{confident}) events only, we include a larger list of ANTARES and ORCA6 triggers and \texttt{PyCBC}~\citep{Usman_2016} offline GW triggers detected during O3. Most of the existing searches rely exclusively on confident triggers to look for a possible coincidence with GWs, overlooking the large population of subthreshold events both in the GW and HEN domains. Incorporating these low-significance triggers is therefore necessary for probing the faint population of joint GW--HEN sources expected near the sensitivity limits of current observatories, motivating the development of dedicated subthreshold, multimessenger frameworks. This search starts by identifying pairs of neutrino and GW triggers that could have a common origin, ranks the pairs according to a statistic accounting for the time and sky proximity between the candidates and their respective significance, and assigns a false alarm rate (FAR [yr$^{-1}$]) to them.
    
The paper is organized as follows: Section~\ref{sec:triggered} describes the methodology and the results of the coherent unmodeled search. Section~\ref{Subthreshold search} presents the data and the method used for the subthreshold search, as well as the associated results. In Section~\ref{population}, we use these results to derive upper limits on the rate densities of BNS and NSBH mergers.

\section{Triggered search}
\label{sec:triggered}

In this section, a triggered search is presented, in which HEN candidates selected from the ANTARES and ORCA6 datasets are used as external triggers to search for GW candidates. As HEN trajectories can be reconstructed with sub-degree resolution~\cite{KM3NeT:2024paj}, using them as external triggers allows for defining narrow GW search regions which can be scanned using a model-independent, computationally-intensive analysis pipeline. The search presented in this section makes use of \texttt{X-Pipeline}~\cite{Sutton_2010}, a software package frequently used in searches for coincidences of weak GW signals with external triggers, including HENs, fast radio bursts \citep{o3afrb}, and GRBs \citep{Abbott2021ApJGRBO3a,LIGOScientific:2021iyk}.

\subsection{Reconstructed event selection}
\label{sec:evtsel}
    
The triggers for the GW search are HEN candidates with a high probability of being astrophysical in origin. As most of the events detected by ANTARES and ORCA6 stem from atmospheric neutrino and muon backgrounds, the corresponding datasets need to be filtered. For both the ANTARES and ORCA6 datasets, we consider only track-like events, as this event topology is associated with the best angular resolution, and muons are rejected by selecting tracks based on their direction and reconstruction quality. 

For ANTARES, we apply the data filtering procedure designed for point-source searches~\cite{ANTARES:2017dda}. In particular, we require events to be upgoing or nearly horizontal, with a zenith angle $\cos\theta>-0.1$ and track reconstruction quality parameter $\Lambda_\nu > -5.2$~\cite{KM3NeT:2023cdr,Fehn:2015ujl}. The reconstructed energy of each event is corrected to account for the ANTARES detector ageing and is required to be larger than \qty{100}{\GeV}. 

For ORCA6, we require events to be strictly upgoing ($\cos\theta>0$) and select them using a Boosted Decision Tree (BDT). This BDT provides a score ($\text{BDT}_{\rm score}$) between $0$ and $1$ that reflects the quality of the reconstruction, with values closer to $0$ ($1$) corresponding to background-like (signal-like) events~\cite{KM3NeT:2023cdr}. We apply a cut on the BDT score, $-\log_{10}(1 - \text{BDT}_{\rm score}) > 4$, and keep only reconstructed events with a reconstructed energy $E_{\rm rec}$ above \qty{1}{\GeV}~\cite{KM3NeT:2023cdr}. 

After these cuts, about 400 (300) events remain in the ANTARES (ORCA6) data, out of which $9\%$ (less than $1\%$) are atmospheric muons according to MC simulations. The distributions as a function of the reconstructed energy and the variable used for the cut are shown in Appendix~\ref{dataset}. For both ANTARES and ORCA6, the uncertainty on the reconstructed energy $E_{\rm rec}$ is of the order of $\delta \log_{10}(E_{\rm rec}/\unit{\GeV}) \sim 0.3$.

Although the selection reduces muon backgrounds~\cite{ANTARES:2017dda,KM3NeT:2023cdr}, the ORCA6 and ANTARES samples remain heavily dominated by atmospheric neutrino events, whose spectrum is softer than that for astrophysical HENs. To evaluate the probability for an event to be of astrophysical origin, we therefore compute a metric called \emph{signalness}, $\mathcal{S}$, initially introduced by the IceCube collaboration~\cite{Abbasi_2023}, defined as 
\begin{equation}
    \mathcal{S}(E_{\rm rec}, X_\nu) = \frac{N_\mathrm{signal}(E_{\rm rec},X_\nu)}{N_\mathrm{signal}(E_{\rm rec},X_\nu) + N_\mathrm{background}(E_{\rm rec},X_\nu)},
    \label{eq:signalness}
\end{equation}
where $N_\mathrm{signal}(E_{\rm rec}, X_\nu)$ and $N_\mathrm{background}(E_{\rm rec},X_\nu)$ are, respectively, the expected numbers of signal and background events with energy $>E_{\rm rec}$ and quality $>X_\nu$. The  $X_\nu$ quantity corresponds to the $\Lambda_\nu$ reconstruction quality parameter for ANTARES and the BDT discriminant for ORCA6. 
For ANTARES, the signalness is computed differently for events with $\cos\theta > 0$ (upgoing) and for events with $-0.1 < \cos\theta < 0$ (downgoing nearly-horizontal). It is computed using only the upgoing events in the former case, while all events with $\cos\theta > -0.1$ are used in the latter. 
Similarly to IceCube~\cite{Abbasi_2023}, we assume flavor equipartition after propagation, with a neutrino flux equal to the best-fit value from the diffuse neutrino flux measured by IceCube~\cite{naab2023measurementastrophysicaldiffuseneutrino}:
\begin{equation}
    \Phi_{\rm 1f}^{\nu+\bar{\nu}} = 1.68\times10^{-18} (E/\qty{100}{\TeV})^{-2.58} \unit{\per\GeV\per\square\cm\per\second\per\steradian}.
    \label{eq:nuflux}
\end{equation}
        
\begin{figure}
    \centering
    \includegraphics[width=0.6\linewidth]{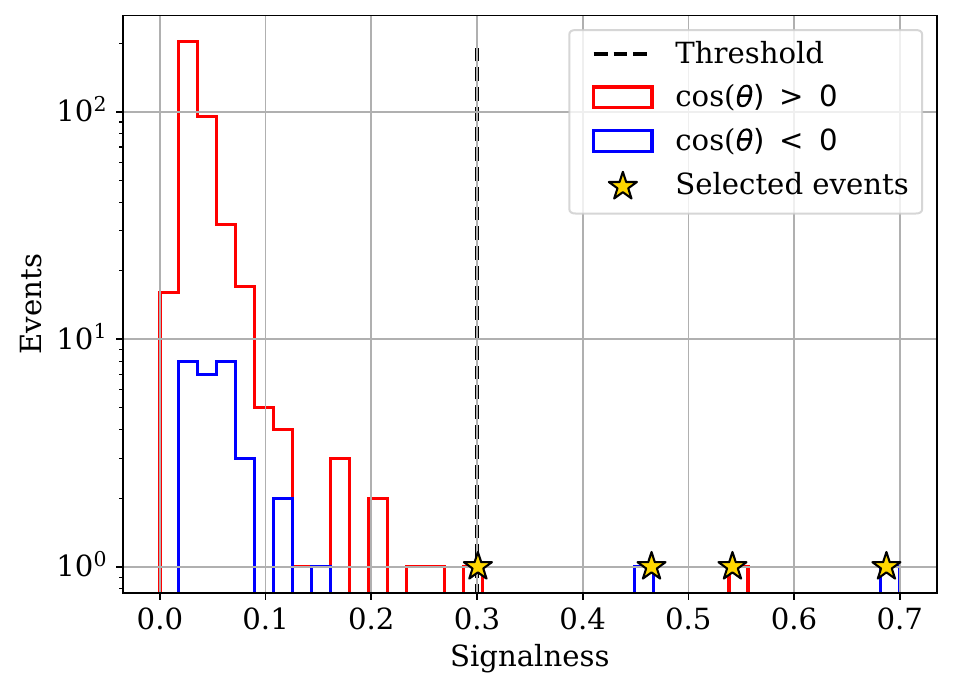}
    \caption{The signalness distribution of ANTARES upgoing reconstructed events (red) and downgoing (nearly horizontal) events (blue). The star markers represent the selected events, whose properties are reported in Table~\ref{tab:ANTARES-selection}.}
    \label{fig:ANTARES-signalness}
\end{figure}

We select reconstructed events with a signalness above $0.3$, again following the criterion introduced by IceCube in \cite{Abbasi_2023}. For ORCA6, no reconstructed event with a signalness above $0.1$ has been found, and, consequently, ORCA6 observations are not used for this triggered search. The ANTARES signalness distributions are shown in Figure~\ref{fig:ANTARES-signalness} in red for upgoing reconstructed events and in blue for downgoing (nearly-horizontal) events. Two upgoing and two downgoing events with a signalness $\mathcal{S} > 0.3$ were identified. Their properties are presented in Table~\ref{tab:ANTARES-selection}.
    
\begin{table*}[]
    \label{tab:ANTARES-selection}
    \centering
     \resizebox{\linewidth}{!}{\begin{tabular}{|c|c|c|c|c|c|c|c|c|c|c|}
        \hline
        {HEN} &
        {Time} &
        {Ra} &
        {Dec} &
        {$E_{\rm rec}$} &
        {$\Lambda_\nu$} & 
        {$\beta$} & 
        {$\cos\theta$} &
        {Signalness $\mathcal{S}$} &
        {Sky localization} & \textit{p}-value\\
        & $[$mjd$]$ & [deg] & [deg] & [TeV] & - & - & - & & error [deg] & GW candidate\\
        \hline
        \#1 & $58735.75391$ & $92.2$ & $44.8$ & $47$ & $-5.10$ & $0.4$ & $0.039$ & $0.30$ & $0.77$ & $0.72$\\
        \#2 & $58793.19922$ & $241.5$ & $27.7$ & $513$ & $-5.02$ & $0.5$ & $-0.0079$ & $0.69$ & $0.54$ & $0.44$\\
        \#3 & $58813.91406$ & $156.4$ & $-21.9$ & $154$ & $-4.88$ & $0.3$ & $0.55$ & $0.54$ & $0.64$ & $0.50$ \\
        \#4 & $58931.10547$ & $59.0$ & $47.4$ & $159$ & $-5.16$ & $0.5$ & $-0.019$ & $0.47$ & $0.64$ & $0.24$\\
        \hline
    \end{tabular}}
        \caption{High-energy astrophysical neutrino candidates selected for the \texttt{X-pipeline} triggered search. This set of events is obtained by applying the point-source selection cuts and the signalness threshold discussed in Section~\ref{sec:evtsel} on ANTARES data collected during O3. Ra and Dec are the reconstructed right ascension and declination, $E_{\rm rec}$ is the reconstructed energy of the event in ANTARES, $\cos\theta$ is the cosine of the zenith angle, $\Lambda_\nu$ is the track reconstruction quality parameter, and $\beta$ is the angular error estimate on the direction. For each event, the sky localization error is defined as the $68^\mathrm{th}$ percentile of the angular separation between the true and reconstructed neutrino directions; its value is extracted from signal simulations and assumed to depend only on the reconstructed energy. For each trigger, the \textit{p}-value of the loudest coincident GW candidate is indicated in the last column (see Section~\ref{config_xpipe}).}
\end{table*}

\subsection{Configuration of the analysis} 
\label{config_xpipe}
    
The four HEN candidates identified in the previous section are used as external triggers to search for a GW counterpart within the GW data using the \texttt{X-pipeline} framework. \texttt{X-pipeline} uses the timings and localization areas of the HEN triggers presented in Table~\ref{tab:ANTARES-selection} to perform a minimally modeled search for GW signals. We considered the three interferometers LIGO Hanford Observatory (LHO), LIGO Livingston Observatory (LLO), and Virgo \citep{o3run}, which took data during O3. Since \texttt{X-pipeline} allows for investigating a broad range of GW waveforms, this analysis probes GW-HEN emission from not only CBCs but also CCSNe \citep{supernova1,supernova2} and accretion disk instabilities (ADI) \citep{adi}. We choose a frequency search window of $20-500$~Hz where the GW interferometers are the most sensitive.

For each HEN, we search for GW events in a $\pm 500$~s on-source time window centered around the HEN trigger. This choice allows for covering the time delays between GW emission and the HEN prompt emission predicted by most models~\cite{2011APH..35..1B}. Backgrounds are estimated using $3$ hours of off-source data collected right before and after the on-source window, and split into intervals of $\qty{1000}{\s}$. The spatial region used for the search is the $2\sigma$ localization area of the HEN, and we present the $1\sigma$ uncertainty in Table~\ref{tab:ANTARES-selection} for the four HEN candidates.

For each HEN trigger, \texttt{X-pipeline} creates a list of GW candidate events in the on-source region by identifying the clusters of pixels with large values of coherent energies, ranks them based on the detection statistic value (summed energies associated with the $h_{\times}$ and $h_+$ polarization channels), selects the highest-ranking candidate, and compares its signal-to-noise ratio (SNR) with the background SNR distribution estimated from off-source regions, as described in~\cite{Sutton_2010}. The \textit{p}-value of a given candidate is then defined as the fraction of background events with an SNR higher than that observed for the candidate~\cite{LIGOScientific:2021iyk}.

If a GW signal overlaps with a loud glitch, it may be misidentified as background, a situation that is particularly likely for long-duration waveforms such as those originating from ADI. In \texttt{X-pipeline}, loud glitches can be mitigated using \emph{autogating}, i.e., zeroing the data from a given interferometer when a large excess above the expected background is observed~\citep{LIGOScientific:2021iyk}. Although this procedure is applied only when no other interferometer records a coherent excess, it can nevertheless reduce the identification efficiency for loud GW transients. We therefore determine whether to apply autogating on a trigger-by-trigger basis; the corresponding criteria are described in what follows.

To estimate the sensitivity of this analysis to a given GW emission model, we evaluate the probability for \texttt{X-pipeline} to recover simulated waveforms injected into off-source data~\citep{Sutton_2010}. The simulated waveforms are based on emission models that predict coincident GW and HEN signals within the on-source time window. In particular, we assume that the hadronic acceleration processes leading to neutrino production take place within the relativistic jet of a (possibly low-luminosity) GRB associated with the GW emitter \cite{GRB_jet1, GRB_jet2, GRB_jet3}. Consequently, we use the same waveform models as in previous combined GW-GRB searches conducted with \texttt{X-pipeline}~\cite{LIGOScientific:2021iyk, o3afrb,Abbott2017ApJ84189A}. The models are drawn from $11$ different sets representing the following three families of models, and the detailed parameters of these injections are provided in Table~\ref{table:waveforms}:
   
\begin{itemize}
    \item \textbf{Sine-Gaussian Chirplet (SGC):} this waveform class is used to simulate GW signals from CCSNe \citep{Szczepanczyk:2023ihe}. We simulate four models with central frequencies $f_0$ of $70, 100, 150$, and $\qty{300}{\hertz}$ and signal durations ranging from $3$ to $\qty{14}{\ms}$, as was done in \cite{Abbott2017ApJ84189A}. We set the energy radiated in GWs to $E_\mathrm{GW} = \qty{e-2}{\solarmass}c^2$. Since the neutrino emission is assumed to occur within a jet, we take the inclination angle $\iota$ of the CCSN rotation axis with respect to the observer to be lower than $\qty{5}{\degree}$~\cite{Abbott2017ApJ84189A}. For each injection, we require this angle to lie between $\qty{0}{\degree}$ and $\qty{5}{\degree}$ and randomly sample it from a uniform distribution in $\cos\iota$.
    \item \textbf{BNS and NSBH inspirals}: these models are used to simulate signals from the coalescence of two neutron stars or of a neutron star and a black hole. For both NSBH and BNS, we sample the neutron star masses from a Gaussian distribution with central value $\qty{1.4}{\solarmass}$ and width $\qty{0.2}{\solarmass}$. For NSBH, we sample the black hole masses from a Gaussian distribution centered on $\qty{10}{\solarmass}$ with width $\qty{6}{\solarmass}$, constraining them to lie in the $\qtyrange{2}{25}{\solarmass}$ range. 
    For both inspiral types, we sample the inclination angles $\iota$ between the total angular momentum vector and the line of sight from a uniform distribution in $\cos\iota$, ranging from $0$ to $\qty{30}{\degree}$~\cite{Abbott2017ApJ84189A}.
    \item \textbf{ADIs}: these waveform models describe the case where a CCSN results in the formation of a rapidly-spinning black hole with a thick accretion torus, turbulence within the torus could lead to GW emission over $\mathcal{O}(10-100)$ seconds~\cite{putten1,putten2}. For this study, we consider five different sets of ADI waveforms based on the van Putten model~\cite{putten1,putten2}, assuming an accretion disk mass of $\qty{1.5}{\solarmass}$, following the approach in~\cite{Abbott2017ApJ84189A}. Their central frequencies, energies, and signal durations are the same as in \cite{Abbott2017ApJ84189A}. 
\end{itemize}
    
For each waveform model, we perform $600$ injections and evaluate the fraction of injections that yield a signal more significant than the loudest GW candidate observed in the data, referred to as \emph{signal efficiency} thereafter. If the signal efficiency is not a monotonic function of the source distance, the presence of a loud glitch in the data is likely. In this case, for the neutrino trigger under consideration, we re-analyze the data using the autogating procedure described above. This is the case for two of the four neutrino triggers studied in this analysis: HEN\#1 and HEN\#3. Finally, for each waveform, we evaluate the \emph{exclusion distance} $D_{90}$ at which the signal efficiency reaches $90$\%.

\begin{table}[h!]
    \centering
    \begin{tabular}{|c|c|c|c|}
    \hline
    {Waveform} &
    {Frequency} &
    {Duration Parameter} & $E_{\rm GW}$\\
    type & [Hz] & [s] & [$M_\odot c^2$]\\
    \hline
    SGC-A & 70 & $14\times10^{-3}$ & $0.01$\\
    SGC-B & 100 & $11\times10^{-3}$ & $0.01$\\
    SGC-C & 150 & $6.9\times10^{-3}$ & $0.01$\\
    SGC-D & 300 & $3.4\times10^{-3}$ & $0.01$\\
    ADI-A & $135-166$ & $38.8$ & $0.02$\\
    ADI-B & $110-209$ & $9.4$ & $0.22$\\
    ADI-C & $130-251$ & $236$ & $0.25$\\
    ADI-D & $119-173$ & $142$ & $0.02$\\
    ADI-E & $111-260$ & $76$ & $0.17$\\
    \hline
    \end{tabular}
    \label{table:waveforms}
    \caption{Frequencies, waveform durations, and energies radiated in GWs for the SGC and ADI models described in Section~\ref{sec:injections}. These parameters are taken to be identical to the ones used in GW-GRB searches in~\cite{Abbott2017ApJ84189A,LIGOScientific:2021iyk}. BNS and NSBH are not shown, as the injection input parameters are different for these sources (see Section~\ref{config_xpipe}).}
\end{table}

\subsection{Results}
\label{resultssection}

Applying \texttt{X-pipeline} to the four HEN triggers identified in Section~\ref{sec:evtsel}, we found no significant coincident GW event. The \textit{p}-values of the most significant GW candidates for the HEN triggers are shown in the last column of Table~\ref{tab:ANTARES-selection}. 

Since no GW signal associated with the HEN triggers was found in this analysis, we estimate exclusion distances using the methodology and models described in Section~\ref{config_xpipe}. The values of $D_{90}$ for each HEN trigger are shown in Table~\ref{table:waveforms_exclusion_distances} for the different waveform models.

\begin{table}[h!]
\centering
\begin{tabular}{|c|S[table-format=3.1]|S[table-format=3.1]|S[table-format=3.1]|S[table-format=3.1]|c|}
    \hline
    {Waveform} & \multicolumn{4}{c|}{$D_{90}$ [Mpc] for each HEN} & $R_\mathrm{UL}$ \\
    \cline{2-5}
    type  & {\#1} & {\#2} & {\#3} & {\#4} & [\unit{\per\cubic\giga\parsec\per\year}] \\
    \hline
    BNS &  22.6 & 47.7 & 36.2 & 72.0 & \num{1.3e6}\\
    NSBH & 49.9 & 106.6 & 93.0 & 157.6 & \num{1.2e5}\\
    SGC-A &  80.4 & 168.5 & 138.6 & 241.7 & \num{2.9e4}\\
    SGC-B &  58.0 & 160.0 & 99.9 & 205.7 & \num{7.6e4}\\
    SGC-C &  28.9 & 112.4 & 52.3 & 149.0 & \num{6.2e5}\\
    SGC-D &  122.2  & 54.8 & 3.6 & 62.7 & \num{3.2e8}\\
    ADI-A & 21.9 & 31.3 & 33.3 & 68.8 & \num{1.4e6}\\
    ADI-B &  80.9 & 138.0 & 140.7 & 212.9 & \num{2.8e4}\\
    ADI-C &  34.6 & 39.8 & 60.7 & 81.9 & \num{3.6e5}\\
    ADI-D &  15.1 & 14.3 & 22.5 & 33.1 & \num{5.1e6}\\
    ADI-E &  35.4 & 46.9 & 68.4 & 72.6 & \num{3.4e5}\\
    \hline
\end{tabular}
\caption{Values of the exclusion distance $D_{90}$ of each waveform used in the analysis for the four HEN analyzed. The last column shows the resulting upper limit on the source rate density, evaluated in Section~\ref{subsec:upperlimits_triggered}.}
\label{table:waveforms_exclusion_distances}
\end{table}

The exclusion distances vary widely depending on the HEN trigger considered because the detectability of GW events strongly depends on the trigger's timing and localization. Moreover, when HEN\#1 and HEN\#3 were observed, only two interferometers, LIGO Livingston and Virgo, were functioning, and consequently, these triggers are associated with the lowest exclusion distances for CBC inspirals and SGC models. 
For ADI waveforms, however, exclusion distances for HEN\#2 are lower than for HEN\#3. Since autogating was not applied for HEN\#2, this result suggests the presence of moderate glitches in the on-source region, which overlap with the long-duration ADI signals. Finally, for each trigger, variations of $D_{90}$ among waveform models reflect the range of waveform parameters considered, as $D_{90}$ scales approximately with the square root of the energy emitted in GWs and with the inverse of the waveform frequency.

Even for the HEN trigger associated with the highest GW sensitivity, the exclusion distances obtained for BNS and NSBH waveforms are significantly smaller than the corresponding horizons of the LIGO/Virgo interferometers. This discrepancy arises because the antenna patterns of the GW detection in the sky regions associated with the different HEN triggers were suboptimal during the time windows considered and because the matched filtering CBC search is more sensitive with respect to the coherent excess energy search for the same type of signals. When the KM3NeT detector approaches completion, future analyses will benefit from a larger number of HEN triggers, some of which are expected to correspond to more favorable GW detection conditions.

\section{Joint subthreshold search}
\label{Subthreshold search}

In this section, a joint search for neutrino and gravitational-wave coincidences, including subthreshold GW and HEN candidates, is presented. In what follows, we first describe the generation of the GW and neutrino data and then present the method used to identify and rank the candidate associations in the joint search.

\subsection{Neutrino and GW data inputs}
\label{subsec:datainput}

Contrary to the triggered search described in Section~\ref{sec:triggered}, this analysis aims at identifying statistically significant coincidences between GW and HEN events which, taken individually, could be considered background-like. The neutrino selection should hence be considerably relaxed while the GW selection should rely on a model-dependent, fast identification algorithm.

\subsubsection{Neutrino event selection}

For ANTARES, selection cuts to reject atmospheric background are similar to the ones presented in Section~\ref{sec:evtsel}, with a slightly relaxed cut on the track reconstruction quality parameter, $\Lambda_\nu > -5.5$. For ORCA6, only reconstructed upgoing events with $-\log_{10}(1 - \text{BDT}_{\rm score}) > 1$ are kept, to remove the bulk of the atmospheric background. Compared with the selection used in Section~\ref{sec:triggered}, the muon contamination rises to 59\% (42\%) for ANTARES (ORCA6), while the astrophysical neutrino selection efficiency, assuming an $E^{-2.5}$ spectrum, increases by a factor of about $1.5$ ($2.5$).

In this work, we do not combine the ANTARES and ORCA6 datasets or analyze them together, as the energy ranges covered and the observables used in the ranking statistic are different.

\subsubsection{Gravitational-wave candidate generation}
\label{gw-generation}

On the GW side, we use the GWTC-3 \citep{o3_dist} released triggers, which correspond to candidate events with a FAR $<$ 2/day, produced by the \texttt{PyCBC} \citep{Usman_2016} search pipeline. The latter is a modeled search for the identification of CBC events in the data. \texttt{PyCBC} produces CBC triggers by finding matches between the data and a grid of template waveforms, the \emph{template bank} \citep{2017arXiv170501845D}, that covers the range of possible source parameters, e.g., masses and spins, and by using matched-filtering techniques \citep{Allen_2012} to discriminate the signal from the background noise. These triggers are submitted to various consistency and coincidence tests \citep{Usman_2016} (such as the $\chi^2$ test) to assign a ranking statistic value to them. Finally, the significance of the resulting triggers represented by their FAR value is computed by comparison with the background. The FAR quantifies how likely it is for noise to produce a trigger with the same or larger ranking statistic as the candidate in question. The \texttt{BAYESTAR} rapid sky localization algorithm \citep{Bayestar, RevModPhys.94.025001} is used to estimate the sky localization of each GW trigger.

\subsection{Ranking statistic}
\label{subsec:ranking_stat}

As introduced in Section~\ref{sec:intro}, we conduct a joint search for coincident GW and neutrino triggers, sorting candidates using a Bayesian ranking statistic inspired by \cite{Ranking}. 
This Bayesian ranking statistic compares the signal to the noise hypothesis given the dataset, therefore it can be written as $\frac{P(D_g, D_\nu| H^C)}{P(D_g, D_\nu| H^N)}$ where $D_g, D_\nu$ are the GW and neutrino dataset, respectively, $H^C$ is the \textit{coincident signal} hypothesis where both GW and neutrino datasets contain noise and signal, and the signals come from the same source, while $H^N$ is the noise hypothesis which can take several forms: (1) $H^{NN}$ hypothesis where both GW and neutrino dataset contain their respective noise only; (2) $H^{NS}$ hypothesis where GW data contain noise and signal, while neutrino data contain only noise; (3) $H^{SN}$ hypothesis where GW data contain noise, and neutrino dataset contains noise and signal; (4) and finally $H^{SS}$ hypothesis: both GW and neutrino dataset contain noise and signal, but the signals are unrelated.
    
As derived in \cite{Ranking}, \cite{SKYOVERLAP}, and \cite{Pillas_2023}, and adapted for the ORCA6 and ANTARES datasets, the Bayesian ranking statistic
\begin{equation}
    \Lambda =\frac{P(D_{g}, D_{\nu}|H^{C})}{P(D_{g}, D_{\nu}|H^{NN} \bigvee H^{SN} \bigvee H^{NS} \bigvee H^{SS})} ~ , 
    \label{eq:ranking-stat1}
\end{equation}
can be simplified as:
\begin{equation}
    \Lambda =\frac{I_{\Delta \mathrm{t}}I_{\Omega}}{1 + Q_{g} + Q_{\nu} + Q_{g}Q_{\nu}},
    \label{eqn:statistics}
\end{equation}
assuming ignorance about prior probabilities between the aforementioned hypotheses. Here, 
\begin{equation}
    Q_{g} = Q_{g}(D_{g}) = \frac{P(D_{g}|\mathrm{noise})}{P(D_{g}|\mathrm{signal})},
\end{equation}
and
\begin{equation}
    Q_\nu = Q_\nu(D_{\nu}) = \frac{P(D_{\nu}|\mathrm{noise})}{P(D_{\nu}|\mathrm{signal})}
    \label{eq:qnu}
\end{equation}
are the single-instrument Bayes factors comparing the noise-only and noise+signal hypotheses in the GW and ANTARES/ORCA6 data, respectively. $I_{\Delta t}$ and $I_\Omega$ quantify the temporal and spatial overlap, respectively. The probability of a GW source being of a given type, e.g., BNS or BBH, is not taken into account, as the number of BNS expected in O3 data is too small for this probability to be reliably estimated.
    
In the following sections, we describe how each term is computed.

\subsubsection{Neutrino trigger significance}
\label{Qnu}

\begin{figure*}
    \centering
    \includegraphics[width=0.48\textwidth]{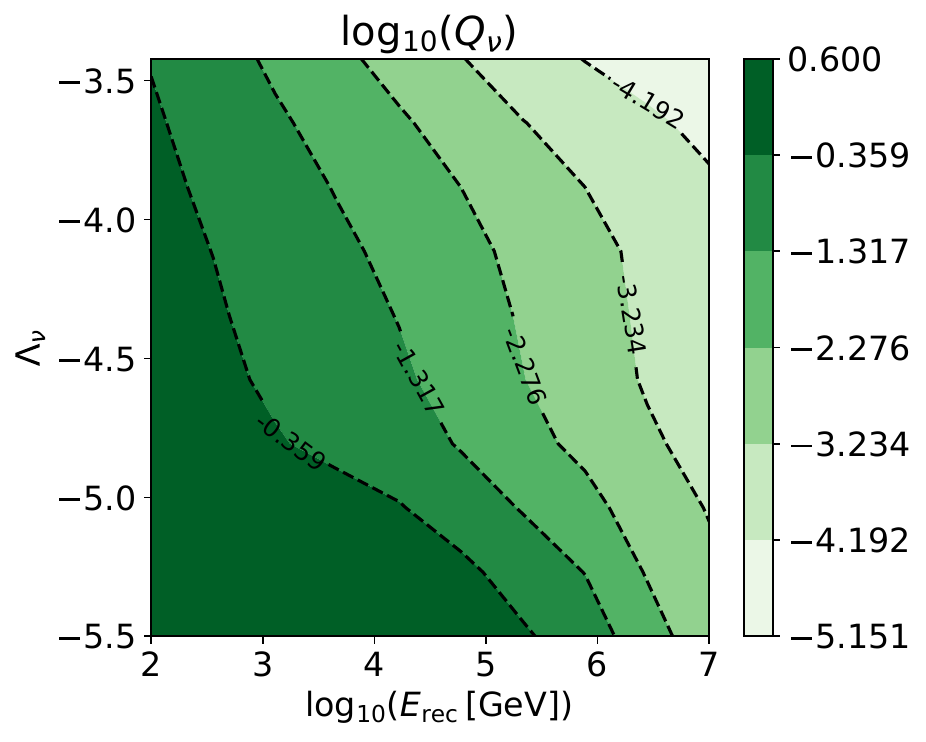}
    \includegraphics[width=0.45\textwidth]{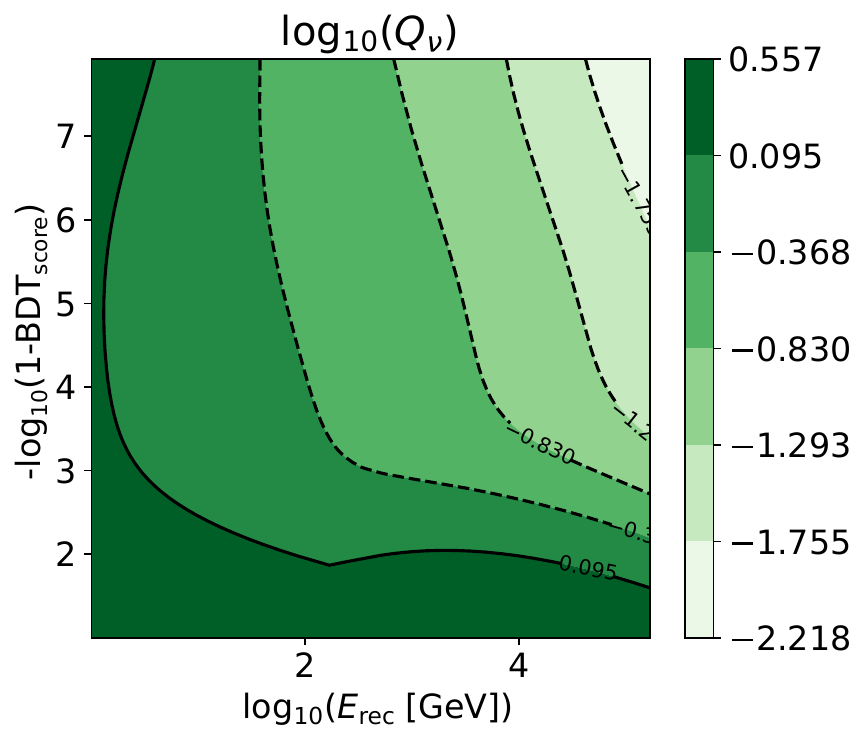}
    \caption{Bayes factor computed with MC astrophysical neutrinos and atmospheric neutrinos and muons as a function of $\log_{10}(E_{\rm rec}\text{ [GeV])}$ and $\Lambda_\nu$ for ANTARES (left), and $\log_{10}(E_{\rm rec}\text{ [GeV])}-$ and $-\log_{10}(1 - \mathrm{BDT}_\mathrm{score})$ for ORCA6 (right).}
    \label{fig:nu-bf}
\end{figure*}

Following the method proposed in  \cite{Pillas_2023} for estimating the Fermi/GBM GRB candidate significance, we build the ORCA6 trigger significance Bayes factor $Q_\nu$ from a Kernel Density Estimation-based method (KDE) \cite{KDE}. We use simulated signals and background to train the KDE in the $\log_{10}$($E_\mathrm{rec}$) -- $X$ observable plane, where $X$ represents either the $\Lambda_\nu$ observable (ANTARES) or $-\log_{10}(1 - \mathrm{BDT}_\mathrm{score})$ (ORCA6). The signal sample is composed of Monte Carlo (MC) astrophysical neutrinos, while the background sample includes atmospheric neutrinos and muons. The three neutrino flavors are included, and we assume here an $E^{-2.5}$ spectrum, which is within the optimal sensitivity windows of both ANTARES and ORCA6. To check the behavior of $Q_\nu$, we first apply the KDE on a random sample. This is shown in Figure~\ref{fig:nu-bf} for both ANTARES (left) and ORCA6 (right). As expected, the Bayes factor takes signal-like values ($\log_{10} Q_\nu < 0$, see Equation~\ref{eq:qnu}) for high $\Lambda_\nu$/BDT score and high reconstructed energy.

\subsubsection{LIGO/Virgo trigger significance}
\label{sec:Qg}

Similarly to \cite{Pillas_2023}, $Q_g$ is defined as the inverse of the Bayes factor coherent versus incoherent ($\mathrm{BCI}$) that compares the probability of being a coherent signal in the entire network versus an incoherent signal and is defined as:
\begin{equation} Q_g^{-1} = \mathrm{BCI} = \frac{P(D_g | C + \mathrm{Gaussian \ noise})}{P(D_g |I + \mathrm{Gaussian \ noise})} ,
\end{equation}
where $C$ and $I$ are respectively the coherent and incoherent signals. The $\mathrm{BCI}$ is provided by \texttt{BAYESTAR} in the maps of the sky localization posterior probability. We expect GW astrophysical signals to have high $\mathrm{log}_{10}(\mathrm{BCI})$ and Gaussian and transient non-Gaussian noise to both have small $\mathrm{log}_{10}(\mathrm{BCI})$.

\subsubsection{Time overlap}
\label{time-overlap}

$I_{\Delta t}$ quantifies the probability for a pair formed by a GW trigger and a HEN trigger to be separated in time by:
\begin{equation}
    \Delta t = t_{\nu} - t_{\mathrm{GW}},
\end{equation}
where $t_{\mathrm{GW}}$ and $t_{\nu}$ are the merger time of the GW candidate and the detection time of the HEN candidate, respectively.
Since we do not have joint GW-HEN detection, this choice of prior is arbitrary. Therefore, we test two different priors: one focused on signals coming from CBC progenitors and a more agnostic one.

In the first case, one of the most probable sources is neutron star mergers producing GWs and a jet in which a fraction of its energy is converted into a short GRB and HENs. We assume the time delay between the short GRB and the neutrino to be close to zero. Similarly to \cite{Pillas_2023, Ranking}, we define the time offset Bayes factor with a triangular shape centered on 0, and we choose an on-source window allowing the neutrino and the GW candidates to be separated by $\Delta t_\mathrm{max} = \qty{30}{\second}$ in time. This prior extends to large time delays, large enough to allow us to detect so-far-unconfirmed phenomena such as short GRB jet precursors \citep{b007b67d637d40e2b52f64012fcf634e, ShatteringFlares, Stachie:2021uky}. 
This first time offset prior can therefore be written as follows:
\begin{equation}
    I_{\Delta \mathrm{t}} = \left\{
    \begin{array}{ll}
        1 - \frac{\vert\Delta \mathrm{t}\vert}{\Delta t_\mathrm{max}} & \mbox{if } |\Delta \mathrm{t}| < \Delta t_\mathrm{max}, \\
        0 & \mbox{otherwise.}
    \end{array}
\right.
\end{equation}
We do not expect the triangular shape of the prior to play a significant role here, and any prior centered on zero and extending down to the on-source window could be used and is expected to give similar results (e.g., a Gaussian prior).

In the second case, we analyze GW and neutrino data within a 1000-second on-source window to cover most of the theoretical models about HEN emission from CBCs (for instance, based on neutrinos emitted from a precursor GRB jet). This value is selected consistently with other neutrino searches targeting GW sources \citep{2011APH..35..1B,IceCube:2022mma,IceCube:2023atb,abbasi2021probingneutrinoemissiongev,ANTARES:2013tra,Albert_2023,KM3NeT:2023cdr}. This second time offset prior is chosen to be a normalized uniform distribution over $t_\mathrm{GW} \pm \Delta t_\mathrm{max}$, with $\Delta t_\mathrm{max} = \qty{500}{\s}$.

\subsubsection{Sky overlap}
\label{sky-overlap}
To quantify how spatially close the GW and neutrino candidates are, we compute a sky overlap Bayes factor (similarly to \citep{SKYOVERLAP}). It is defined as
\begin{equation}
    I_{\Omega} = \int_{}^{} \frac{P(\Omega | D_g)P(\Omega | D_\nu)}{P(\Omega)} d\Omega
    \label{Iomega}
\end{equation}
where $P(\Omega \vert D_X)$ represents the probability density distribution of the GW ($X=g$) or the neutrino ($X=\nu$) localization, expressed as a function of the solid angle $\Omega$, given the data $D_X$, and $P(\Omega)$ is the prior on the source localization, here equal to $1/4\pi$\,\unit{\per\steradian}. Large values of $I_{\Omega}$ (strongly positive) are expected when the GW and neutrino triggers are overlapping and well localized. $I_{\Omega}$ close to 0  means that the overlap is small, and $I_{\Omega} \approx 1$ indicates that although the two triggers are overlapping, their skymaps are uninformative (large sky localization error box on both sides). As mentioned in Section~\ref{gw-generation}, the sky localization of the GW triggers is given as a skymap computed with \texttt{BAYESTAR}.

\begin{figure*}
    \centering
    \includegraphics[width=0.45\textwidth]{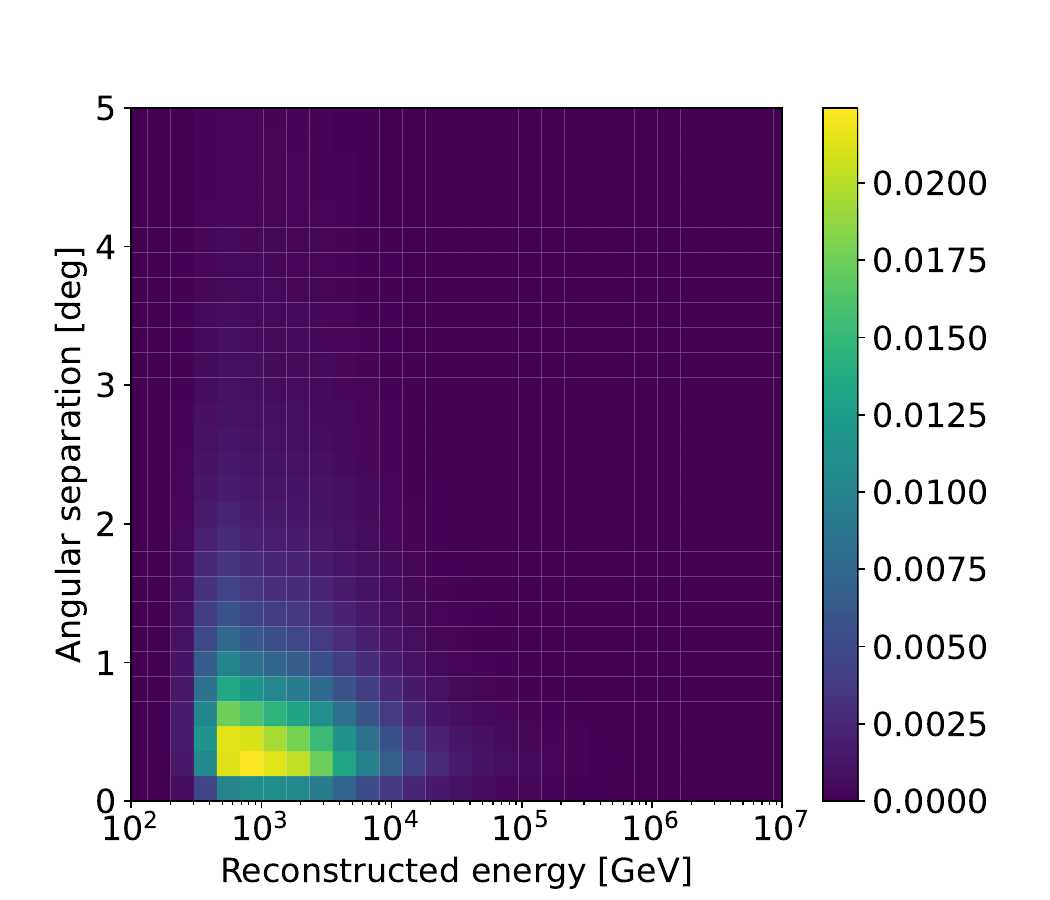}
    \includegraphics[width=0.45\textwidth]{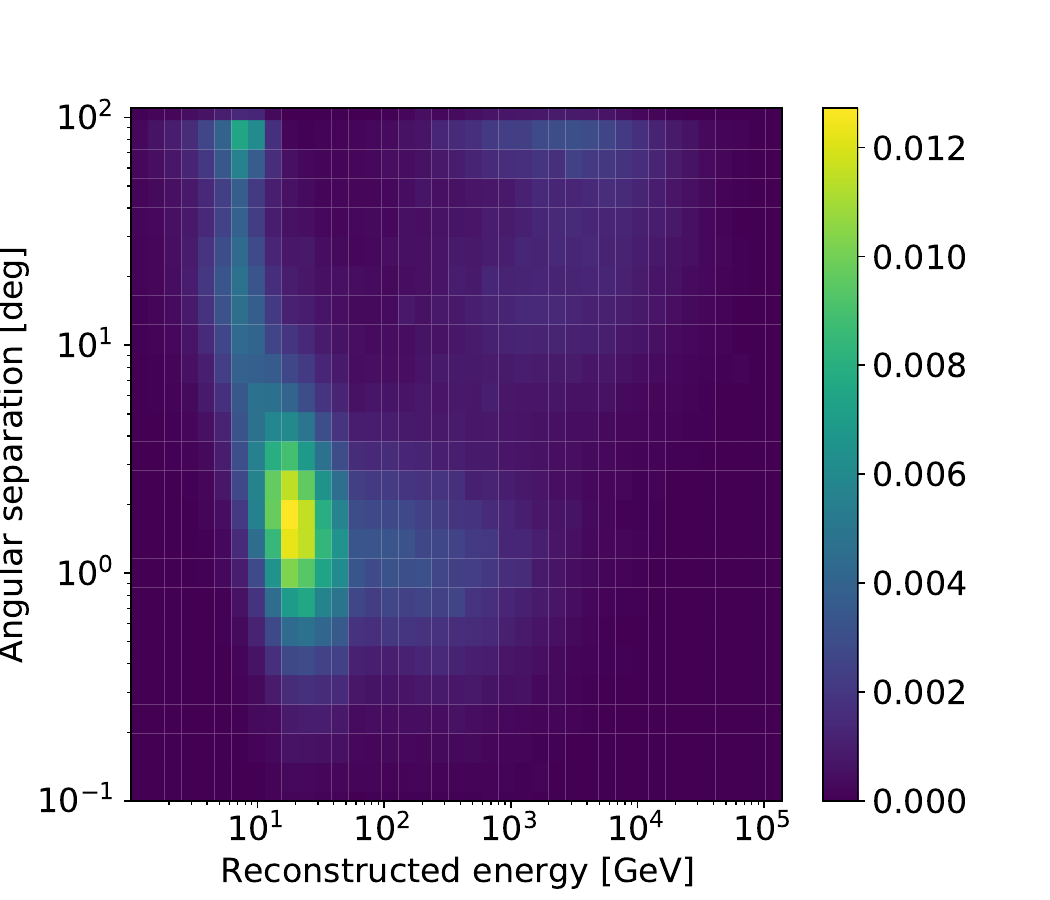}
    \caption{Normalized distribution of the angular separation between the true and reconstructed sky location of the MC astrophysical neutrino sample for ANTARES (left) and ORCA6 (right) as a function of the reconstructed energy.}
    \label{fig:angular-sep-MC}
\end{figure*}

The neutrino sky localization probability $P(\Omega | D_\nu)$ is estimated from MC simulations as a function of energy. It is modeled as a two-dimensional Gaussian whose standard deviation is the 68$^\mathrm{th}$ percentile of the distribution (therefore slightly larger than 1$\sigma$), for each bin of energy, of the angular separation between the true neutrino direction and the reconstructed one, as illustrated in Figure~\ref{fig:angular-sep-MC}. For ORCA6, due to the limited size of the detector, the angular resolution worsens for energies above $\qty{1}{\TeV}$, therefore producing simulated events with a reduced value of $I_\Omega$. As $I_\Omega$ and the neutrino significance $Q_\nu$ are built upon the reconstructed energy, correlations between the two Bayes factors can be introduced. However, these correlations are negligible according to Figures~\ref{fig:corner-injections} and \ref{fig:corner-injections-500s} in Appendix~\ref{sec:limitations}.

\subsection{Modeling signal and background associations}
\label{subsec:model_associations}

To compute the significance of each foreground association candidate, represented by the FAR [yr$^{-1}$] or Inverse FAR (IFAR [yr]), we compare them to a background sample. The FAR is assigned by calculating the rate of background with a ranking statistic equal to or higher than the one observed for the foreground association in question. In what follows, we describe how we model background associations and how we validate the choice of ranking statistic and the FAR evaluation using simulated foreground associations.

\subsubsection{Background associations}
\label{sec:background-results}

\begin{figure*}
    \centering
    \includegraphics[width=0.49\textwidth]{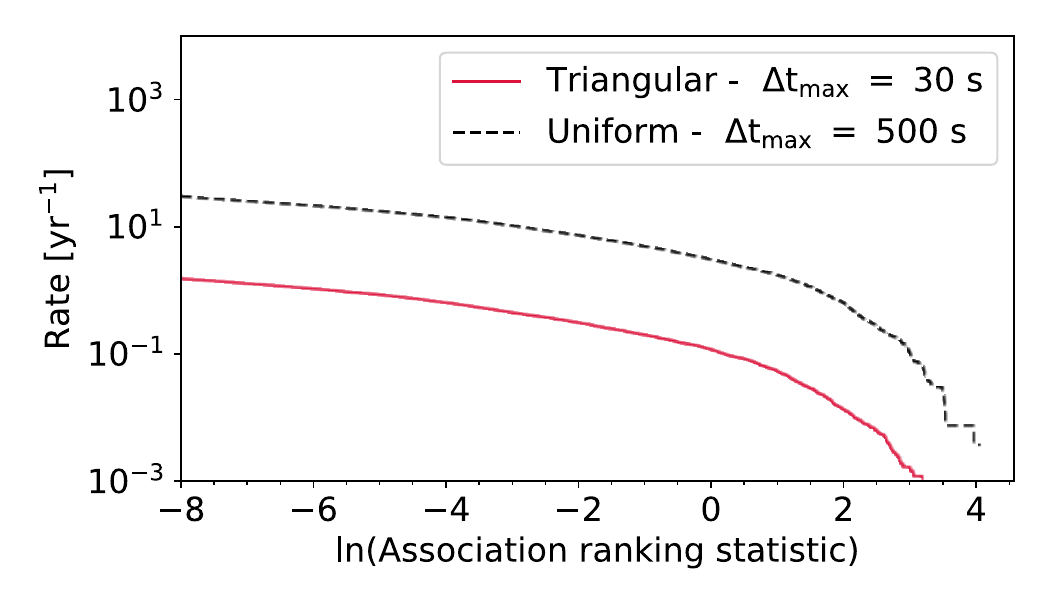}
    \includegraphics[width=0.49\textwidth]{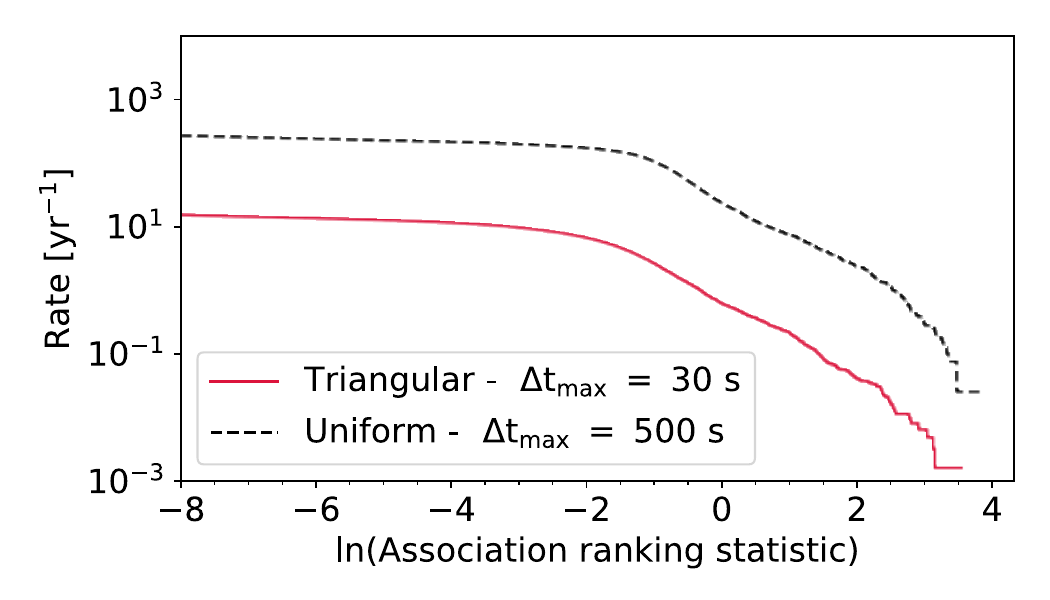}
    \caption{Background rate as a function of the ranking statistic, with 30 s maximum time offset (red solid line) and 500 s maximum time offset (black dashed line) for ANTARES (left) and ORCA6 (right).}
    \label{fig:background-plot}
\end{figure*}

Background associations are generated following a standard method in GW data analysis (e.g., in \textbf{PyCBC}~\citep{W_s_2009}), in which we shift in time the neutrino triggers by a chosen offset (larger than the on-source time window), and we look for coincidences between the GW and the time-shifted neutrino triggers. We repeat this process for several time shifts in order to accumulate the background. To keep the detectors in a local state and to avoid high computational cost, the background interval should be large enough to maximize the significance of the foreground. Here we chose a time interval of $\pm \qty{200000}{\s}$ in steps of \qty{70}{\s} and \qty{1100}{\s}, for the $\pm \qty{30}{\second}$ and the $\pm \qty{500}{\second}$ on-source time window, respectively. The time shifts of the neutrino triggers result in a shift of their sky localization according to the Earth's rotation.

In Figure~\ref{fig:background-plot}, the behavior of the ranking statistic is shown for the time-shifted associations, applying the analysis to ANTARES (left panel) and ORCA6 (right panel) separately. The association ranking statistic $\Lambda$ is less than $1$ for most of the background associations, with the event rate sharply dropping for $\ln\Lambda \gtrsim 2$ in both the GW-ANTARES and GW-ORCA6 analyses. The maximum background association ranking statistic is $\Lambda \sim 59$.
\begin{landscape}
\begin{table*}
\centering
\resizebox{\linewidth}{!}{\begin{tabular}{|c|c|c|c|c|c|c|c|c|c|c|c|c|c|c|c|c|} 
    \multicolumn{14}{c}{a) ANTARES}\\
    \hline
    \multicolumn{1}{|c}{} & \multicolumn{4}{|c}{GW Properties} & \multicolumn{3}{|c}{Neutrino Properties} & \multicolumn{8}{|c|}{Joint Properties} \\
    \hline
    Rank & Merger & Source & $M_{\rm tot, bank}$ & $Q_g$ & $E_{\rm rec}$ & $\Lambda_\nu$  & $Q_{\nu}$ & \multicolumn{2}{c|}{$\Delta t$ $[s]$} & \multicolumn{2}{c|}{$I_{\Omega}$} & \multicolumn{2}{c|}{Time shift $[s]$}  & \multicolumn{2}{c|}{$\Lambda$} \\ \cline{9-16}
    & Time & class. & [$M_\odot$] & & [GeV] & & & \multicolumn{2}{c|}{for $\Delta t_\mathrm{max}$} & \multicolumn{2}{c|}{for $\Delta t_\mathrm{max}$} & \multicolumn{2}{c|}{for $\Delta t_\mathrm{max}$} & \multicolumn{2}{c|}{for $\Delta t_\mathrm{max}$} \\ 
    &  &&  & & &  & & \multicolumn{1}{c}{\qty{30}{\second}} & \multicolumn{1}{c|}{\qty{500}{\second}} & \multicolumn{1}{c}{\qty{30}{\second}} & \multicolumn{1}{c|}{\qty{500}{\second}} & \multicolumn{1}{c}{\qty{30}{\second}} & \multicolumn{1}{c|}{\qty{500}{\second}} & \multicolumn{1}{c}{\qty{30}{\second}} & \multicolumn{1}{c|}{\qty{500}{\second}} \\ 
    \hline
     1 & $1248184675.03$ & Terr $\approx 1$ & $2.1$ & $6.4\times10^{-3}$  &  $7.8\times10^{3}$ & $-4.2$ & $6.1\times10^{-2}$ & $7.75$ & $217.75$ & $56.2$ & $62.7$ & $60390$ & $60600$ & $39.0$ & $58.7$ \\
     2 & $1241719652.42$  & Terr $\approx 1$ & $43.6$ & $4.2\times10^{-4}$ & $4.2\times10^{2}$ & $-4.3$ & $4.7\times10^{-1}$ & $5.146$ & $-384.854$ & $52.6$ & $77.0$ & $101690$ & $101300$  & $29.5$ & $52.1$ \\
     \hline
\end{tabular}}
 
\vspace{0.5cm}
\resizebox{\linewidth}{!}{\begin{tabular}{|c|c|c|c|c|c|c|c|c|c|c|c|c|c|c|c|c|} 
    \multicolumn{14}{c}{b) ORCA6}\\
    \hline
    \multicolumn{1}{|c}{} & \multicolumn{4}{|c}{GW Properties} & \multicolumn{3}{|c}{Neutrino Properties} & \multicolumn{8}{|c|}{Joint Properties} \\
    \hline
    Rank & Merger & Source & $M_{\rm tot, bank}$ & $Q_g$ & $E_{\rm rec}$ & BDT  & $Q_{\nu}$ & \multicolumn{2}{c|}{$\Delta t$ [s]} & \multicolumn{2}{c|}{$I_{\Omega}$} & \multicolumn{2}{c|}{Time shift $[s]$}  & \multicolumn{2}{c|}{$\Lambda$} \\ \cline{9-16}
    & Time & class. & [$M_\odot$] & & [GeV] & score & & \multicolumn{2}{c|}{for $\Delta t_\mathrm{max}$} & \multicolumn{2}{c|}{for $\Delta t_\mathrm{max}$} & \multicolumn{2}{c|}{for $\Delta t_\mathrm{max}$} & \multicolumn{2}{c|}{for $\Delta t_\mathrm{max}$} \\  
    &  &&  & & &  & & \multicolumn{1}{c}{\qty{30}{\second}} & \multicolumn{1}{c|}{\qty{500}{\second}} & \multicolumn{1}{c}{\qty{30}{\second}} & \multicolumn{1}{c|}{\qty{500}{\second}} & \multicolumn{1}{c}{\qty{30}{\second}} & \multicolumn{1}{c|}{\qty{500}{\second}} & \multicolumn{1}{c}{\qty{30}{\second}} & \multicolumn{1}{c|}{\qty{500}{\second}} \\
    \hline
    1 & $1268747361.18$ & Terr $\approx 1$ & $63.9$ & $4.3\times 10^{-1}$ & $20.4$ & $4.6$ & $6.1\times10^{-2}$ & $7.82$ & $257.820$ & $109.1$ & $102.0$ & $96650$ & $96900$ & $53.3$ & $34.1$ \\
    2 & $1266807848.74$ & Terr $\approx 1$ & $9.1$ & $3.4\times 10^{-2}$ & $57.6$ & $5.8$ & $1.4\times10^{-2}$ & $-8.64$ & $281.360$ & $47.9$ & $45.5$ & $-13390$ & $-13100$ & $32.5$ & $32.1$ \\
    \hline
 \end{tabular}}
 
  \caption{Properties of the first two highest-ranked background associations with a) ANTARES and b) ORCA6. $\Delta t_\mathrm{max}$: maximum time delay (either 30 s or 500 s). $M_{\rm tot, bank}$: total mass corresponding to the best-fit waveform in PyCBC template bank. $Q_g$: GW Bayes Factor. $E_{\rm rec}$: reconstructed energy of the event in ANTARES or ORCA6. $Q_{\nu}$: Neutrino Bayes Factor. $\Delta t$: time delay between the neutrino and the GW trigger. $I_{\Omega}$: sky overlap value. $\Lambda$: association rank value. \emph{Source class.} is the GW source classification according to the LVK analysis that can be either BBH, NSBH, BNS, or Terrestrial (noted Terr) here.}
 \label{table:background-assoc}
\end{table*}
\end{landscape}

The properties of the highest-ranked background associations for both analyses are presented in  Table~\ref{table:background-assoc}. The highest-ranked background associations are the same in the $t_\mathrm{max}=\qty{30}{\second}$ and $t_\mathrm{max}=\qty{500}{\second}$ searches, which is expected as the same set of triggers has been used. For these more significant background associations, the sky localization areas of the neutrino and GW triggers have a large overlap; the neutrino trigger is signal-like; and the GW candidate is classified as noise.

\subsubsection{Foreground associations: injection study}
\label{sec:injections}

To verify the validity of the method described in Sections~\ref{subsec:datainput} and \ref{subsec:ranking_stat} and investigate the performance of the search, simulated GW-HEN associations are injected into the data. These associations are composed of MC astrophysical neutrinos and of GW injected signals, generated close in time and at the same sky location. We then estimate the fraction of recovered injections as a function of the IFAR. 

To generate foreground associations, we choose $300$ MC astrophysical neutrinos, with an $E^{-2.5}$ spectrum: $100$ for each merger type (BNS, NSBH, and BBH). For this performance study, we focus on well-reconstructed events, and therefore only select simulated neutrino events with an angular separation between the true and reconstructed sky localizations lower than $\qty{5}{\degree}$, as we aim to measure the ranking efficiency conditional on trigger generation and the MC neutrino pre-selection.
For each simulated neutrino, we inject a GW simulated signal using the priors defined in Section~\ref{time-overlap}, and at the same sky localization. The other GW properties are sampled uniformly from the interval described in Table~\ref{tab:injection-distribution}. The black-hole mass is restricted to lie in the $m_1 \in [3,20] \unit{\solarmass}$ range, as we assumed that NSBH mergers with a very large mass ratio $q = \frac{m_1}{m_2}$ do not result in a GRB jet production and are hence less likely to be associated with neutrino production~\cite{Foucart:2018rjc}. Then the skymaps are generated with \texttt{BAYESTAR} for O3 data, using the same template for simulating the signal and localizing it. 
Removing cases where no GW detector was in observing mode or where the SNR of the injection is below $5$ in all the available detectors (PyCBC trigger not produced), we finally obtain a sample of $144$ BBH-, NSBH- and BNS-like simulated associations for both the $\Delta t_\mathrm{max}=\qty{30}{\second}$ and $\Delta t_\mathrm{max}=\qty{500}{\second}$ searches with ANTARES and $126$ ($153$) for the $\Delta t_\mathrm{max}=\qty{30}{\second}$ ($\Delta t_\mathrm{max}=\qty{500}{\second}$) searches with ORCA6. 

\begin{table*}
    \centering
    \resizebox{\textwidth}{!}{\begin{tabular}{|c|c|c|c|c|c|c|c|c|}
        \hline
        Merger & $m_1$ & $m_2$ & $s_{1,z}$ & $s_{2,z}$ & Inclination  & Phase  & Polarization  & Distance \\
        type & [$M_\odot$] & [$M_\odot$] &  & & [deg] &  [deg] &  [deg] &  [Mpc]\\
        \hline
        BNS & $[1,3]$ & $[1,3]$ & $[-0.1,0.1]$ & $[-0.1,0.1]$ & $\arccos[-1,1]$ & $[0,2\pi]$ & $[0,\pi]$ & $[40,300]$ \\
        (\texttt{TaylorF2}) & & & & & & & & \\
        \hline
        NSBH & $[3,20]$ & $[1,3]$ & $[-0.9,0.9]$ & $[-0.1,0.1]$ & $\arccos[-1,1]$ & $[0,2\pi]$ & $[0,\pi]$ & $[40,300]$ \\
        (\texttt{SEOBNRv4}) & & & & & & & & \\
        \hline
        BBH & $[3,100]$ & $[3,100]$ & $[-0.9,0.9]$ & $[-0.9,0.9]$ & $\arccos[-1,1]$ & $[0,2\pi]$ & $[0,\pi]$ & $[40,1000]$ \\
        (\texttt{SEOBNRv4}) & & & & & & & & \\
        \hline
    \end{tabular}}
    \caption{GW injection parameter ranges used to generate the simulated skymaps with \texttt{BAYESTAR}. These distributions are chosen to be uniform across these intervals. The symbols $s_{1,2z}, m_{1,2}$ denote the spins of the compact objects and their masses in the source frame.
    }
    \label{tab:injection-distribution}
\end{table*}

\begin{figure*}
    \centering
    \includegraphics[width=0.49\textwidth]{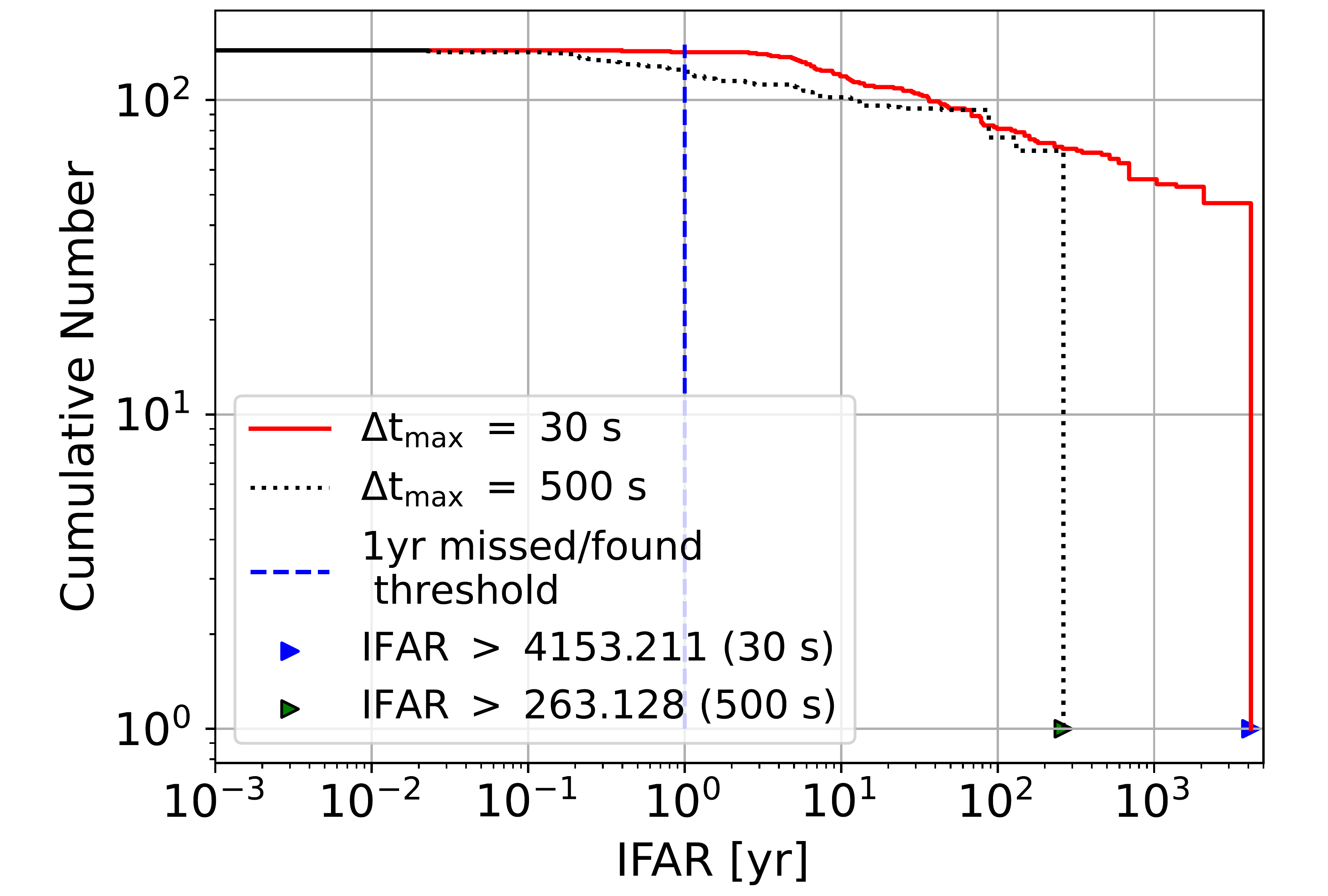}
    \includegraphics[width=0.49\textwidth]{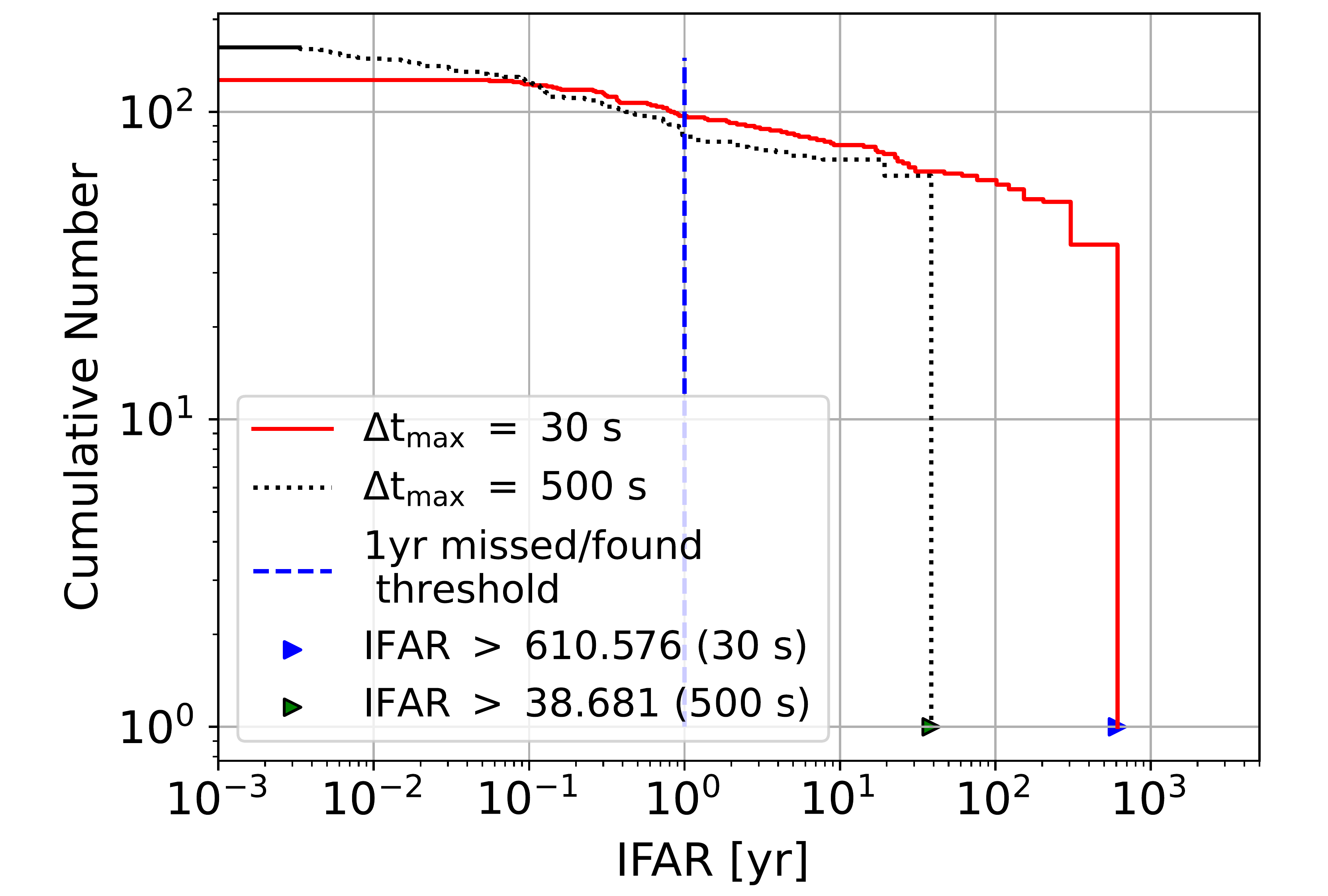}
    \caption{Cumulative number of simulated associations as a function of their IFAR [yr] for the $\Delta t_\mathrm{max} = \qty{30}{\second}$ (red-solid line) and the $\Delta t_\mathrm{max} = \qty{500}{\second}$ analysis (black-dotted line). Results obtained with ANTARES (left) and ORCA6 (right) data.}
    \label{fig:injections-IFAR}
\end{figure*}

To estimate the significance of the simulated associations, an \emph{exclusive} FAR is computed by counting the number of
background associations from Section~\ref{sec:background-results} whose ranking statistic is higher than the one for the simulated association considered. This number is called an \emph{exclusive} FAR as we do not include the simulated signals in the background computation and, hence, this method sometimes leads to slightly higher IFAR values than with an inclusive FAR computation. The cumulative number of simulated associations as a function of the IFAR is shown in Figure~\ref{fig:injections-IFAR} for the two search windows. Here, the associations with an IFAR larger than $\qty{1}{\year}$ are considered to be confidently recovered. For the most significant foreground associations, which rank higher than all the background associations in our sample, only lower limits on the IFAR can be derived. These lower limits are IFAR $> \qty{4153}{\year}$ for the $\Delta t_\mathrm{max} = \qty{30}{\second}$ search and $> \qty{263}{\year}$ for the $\Delta t_\mathrm{max} = \qty{500}{\second}$ search with ANTARES. The respective values for ORCA6 are $> \qty{610}{\year}$ and $> \qty{38}{\year}$. The lower value of the maximum IFAR for the $\pm \qty{500}{\second}$ search with respect to the $\pm \qty{30}{\second}$ search is due to the smaller number of background associations generated.

We show the most significant simulated associations for both searches with ANTARES and ORCA6 in Table~\ref{table:ANTARES-ORCA6-simulated-assoc}. The top associations contain significant GW injections with small $Q_g$ (signal-like) and MC neutrinos with a small sky localization uncertainty and a large overlap (with very large $I_\Omega$). Overall, while a few associations, studied in detail in Appendix~\ref{sec:limitations}, are associated with an IFAR lower than $\qty{1}{\year}$, most of the simulated associations are well recovered. In what follows, we therefore apply our methodology to LIGO-Virgo observations and to data from either ANTARES or ORCA6, and describe the resulting foreground associations.

\begin{landscape}
\begin{table*}
\centering
\resizebox{\linewidth}{!}{\begin{tabular}{|c|c|c|c|c|c|c|c|c|c|c|c|c|}
    \multicolumn{13}{c}{a) ANTARES}\\
    \hline
    & \multicolumn{1}{c}{} & \multicolumn{4}{|c}{GW Properties} & \multicolumn{3}{|c}{Neutrino Properties} & \multicolumn{4}{|c|}{Joint Properties} \\
    \hline
    $\Delta t_\mathrm{max}$ & Rank & Merger & Coherent & $M_{\rm tot, bank}$ & $Q_g$ & $E_{\rm rec}$ & $\Lambda_\nu$ & $Q_{\nu}$ & $\Delta t$ & $I_{\Omega}$  & $\Lambda$ & IFAR \\
    {[s]} & & Time & SNR & [$M_\odot$] & & [GeV] & & & [s] & & & [yr] \\ 
    \hline
    \multirow{3}{*}{30} & $1$ & $1267945365.89$ & $76.7$ & $94.1$ & $8.7\times10^{-14}$ & $8.2\times10^{5}$ & $-5.1$ & $1.6\times10^{-2}$ & $12.24$ & $1387.1$ & $808.4$ & $>4153.2$ \\
    & $2$ & $1269348383.31$ & $183.0$ & $127.2$ & $6.5\times10^{-15}$ & $2.5\times10^{5}$ & $-5.0$ & $3.1\times10^{-2}$ & $-16.93$ & $1167.9$ & $493.2$ & $>4153.2$ \\
    & $3$ & $1258651716.98$ & $148.3$ & $188.4$ & $2.7\times10^{-15}$ & $1.4\times10^{6}$ & $-4.5$ & $7.5\times10^{-4}$ & $-19.65$ & $1389.9$ & $479.1$ & $>4153.2$ \\
    \hline
    \multirow{3}{*}{500} & $1$ & $1269348133.01$ & $196.1$ & $121.0$ & $1.4\times10^{-11}$ & $2.4\times10^{6}$ & $-5.0$ & $2.3\times10^{-2}$ & $233.37$ & $1906.5$ & $1863.1$ & $>263.1$ \\
    & $2$ & $1268087946.35$ & $49.7$ & $52.7$ & $1.9\times10^{-11}$ & $2.2\times10^{6}$ & $-4.9$ & $1.0\times10^{-3}$ & $-79.07$ & $959.1$ & $958.1$ & $>263.1$ \\
    & $3$ & $1264648616.51$ & $179.7$ & $88.0$ & $7.5\times10^{-13}$ & $1.9\times10^{6}$ & $-5.0$ & $1.3\times10^{-3}$ & $6.32$ & $881.8$ & $880.9$ & $>263.1$ \\ [1ex] 
     \hline
\end{tabular}}
\par
\vspace{0.5cm}
 
\resizebox{\linewidth}{!}{\begin{tabular}{|c |c |c |c |c |c |c |c |c |c |c |c |c |}
\multicolumn{13}{c}{b) ORCA6}\\
    \hline
    & \multicolumn{1}{c}{} & \multicolumn{4}{|c}{GW Properties} & \multicolumn{3}{|c}{Neutrino Properties} & \multicolumn{4}{|c|}{Joint Properties}\\
    \hline
    $\Delta t_\mathrm{max}$ & Rank & Merger & Coherent & $M_{\rm tot, bank}$ & $Q_g$ & $E_{\rm rec}$ & BDT & $Q_{\nu}$ & $\Delta t$ & $I_{\Omega}$  & $\Lambda$ & IFAR \\
    $[s]$ & & Time & SNR & [$M_\odot$] & & [GeV] & score & & $[s]$ & & & [yr] \\
    \hline
    \multirow{3}{*}{30} & $1$ & $1265601955.16$ & $13.4$ & $94.3$ & $1.2\times10^{-5}$ & $88.8$ & $5.3$ & $3.5\times10^{-1}$ & $0.34$ & $338.3$ & $248.4$ & $>610.6$ \\
    & $2$ & $1268911385.63$ & $179.0 $ & $107.5$ & $2.0\times10^{-18}$ & $34.0$ & $4.3$ & $5.7\times10^{-1}$ & $2.72 $ & $424.0$ & $245.4$ & $>610.6$ \\
    & $3$ & $1266637630.25$ & $10.9$ & $12.8$ & $2.1\times10^{-5}$ & $45.5$ & $4.5$ & $5.0\times10^{-1}$ & $2.69$ & $337.3$ & $204.2$ & $>610.6$ \\
    \hline
    \multirow{3}{*}{500} & $1$ & $1267737597.94$ & $13.6$ & $12.2$ & $4.8\times10^{-5}$ & $131.6$ & $3.5$ & $5.2\times10^{-1}$ & $398.46$ & $411.2$ & $269.6$ & $>38.7$ \\
    & $2$ & $1267734153.12$ & $17.3$ & $21.4$ & $2.5\times10^{-7}$ & $59.4$ & $4.8$ & $5.1\times10^{-1}$ & $-447.97$ & $382.2$ & $253.4$ & $>38.7$ \\
    & $3$ & $1268572668.02$ & $20.7$ & $20.7$ & $1.1\times10^{-3}$ & $45.5$ & $5.0$ & $5.4\times10^{-1}$ & $-1.75$ & $337.3$ & $218.5$ & $>38.7$ \\ [1ex] 
    \hline
\end{tabular}}

\caption{Properties of the first three most significant simulated associations with a) ANTARES and b) ORCA6. $\Delta t_\mathrm{max}$: maximum time delay (either \qty{30}{\second} or \qty{500}{\second}). $M_{\rm tot, bank}$: total mass extracted from the corresponding waveform in PyCBC template bank. $Q_g$: GW Bayes Factor. $E_{\rm rec}$: reconstructed energy of the MC neutrino. $Q_{\nu}$: Neutrino Bayes Factor. $\Delta t$: time delay between the neutrino and the GW trigger. $I_{\Omega}$: sky overlap value. $\Lambda$: association rank value. IFAR: Joint Inverse False Alarm Rate [yr]. The coherent SNR is the signal-to-noise ratio for the full network, defined as the quadratic sum of the SNRs of the different interferometers.}
 \label{table:ANTARES-ORCA6-simulated-assoc}
\end{table*}
\end{landscape}

\begin{landscape}
\begin{table*}
\centering
\resizebox{\linewidth}{!}{\begin{tabular}{|c |c |c |c |c |c |c |c |c |c |c |c |c |}
    \multicolumn{13}{c}{a) ANTARES}\\
    \hline
    & \multicolumn{1}{c}{} & \multicolumn{4}{|c}{GW Properties} & \multicolumn{3}{|c}{Neutrino Properties} & \multicolumn{4}{|c|}{Joint Properties} \\
    \hline
    $\Delta t_\mathrm{max}$ & Rank & Merger & Source & $M_{\rm tot, bank}$ & $Q_g$ & $E_{\rm rec}$ & $\Lambda_\nu$ & $Q_{\nu}$ & $\Delta t$ & $I_{\Omega}$  & $\Lambda$ & IFAR \\
    $[s]$ & & Time & class. & [$M_\odot$] & & [GeV] & score & & $[s]$ & & & [yr] \\
    \hline
    \multirow{3}{*}{30} & $1$ & $1239317205.14 $ & Terr $\approx 1$ & $8.9$ & $7.3\times10^{-2}$ & $1876.24$ & $-5.5$ & $2.37$ & $25.46$ & $2.5\times10^{-2}$ & $1.0\times10^{-3}$ & $0.8$ \\
    & $2$ & $1268903511.32$ & BBH $\approx 0.62$ & $71.3$ & $6.3\times10^{-4}$ & $2414.5$ & $-5.3$ & $2.18$ & $-18.53$ & $5.2\times10^{-3}$ & $6.2\times10^{-4}$ & $0.7$ \\
    & $3$ & $1258857471.59$ & Terr $\approx 1$ & $14.2$ & $2.0\times10^{-2}$ & $922.1$ & $-5.0$ & $8.0\times10^{-1}$ & $-3.16$ & $7.7\times10^{-5}$ & $3.8\times10^{-5}$ & $0.5$ \\
    \hline
    \multirow{3}{*}{500} & $1$ & $1260264650.62$ & Terr $\approx 1$ & $15.7$ & $1.3\times10^{-2}$ & $668.9$ & $-5.0$ & $9.1\times10^{-1}$ & $-483.05$ & $2.86$ & $1.48$ & $0.4$ \\
    & $2$ & $1239808822.38$ & Terr $\approx 1$ & $3.3$ & $3.4\times10^{-2}$ & $2054.5$ & $-4.9$ & $4.9\times10^{-1}$ & $145.19$ & $1.65$ & $1.1$ & $0.3$ \\
    & $3$ & $1261839789.18$ & Terr $\approx 1$ & $5.5$ & $1.5\times10^{-1}$ & $13060.7$ & $-4.8$ & $1.6\times10^{-1}$ & $-171.18$ & $3.8\times10^{-1}$ & $2.8\times10^{-1}$ & $0.2$ \\
    \hline
\end{tabular}}
 
\vspace{0.5cm}
 
\resizebox{\linewidth}{!}{\begin{tabular}{|c |c |c |c |c |c |c |c |c |c |c |c |c |}
    \multicolumn{13}{c}{b) ORCA6}\\
    \hline
    & \multicolumn{1}{c}{} & \multicolumn{4}{|c}{GW Properties} & \multicolumn{3}{|c}{Neutrino Properties} & \multicolumn{4}{|c|}{Joint Properties} \\
    \hline
    $\Delta t_\mathrm{max}$ & Rank & Merger & Source & $M_{\rm tot, bank}$ & $Q_g$ & $E_{\rm rec}$ & BDT & $Q_{\nu}$ & $\Delta t$ & $I_{\Omega}$  & $\Lambda$ & IFAR \\
    $[s]$ & & Time & class. & [$M_\odot$] & & [GeV] & score & & $[s]$ & & & [yr] \\ 
    \hline
    \multirow{2}{*}{30} & $1$ & $1266460242.30$ & Terr $\approx 1$ & $5.15$ & $6.4\times10^{-4}$ & $6271.4$ & $2.2$ & $7.0\times10^{-1}$ & $-16.70$ & $9.2\times10^{-1}$ & $2.4\times10^{-1}$ & $0.2$ \\
    & $2$ & $1267957899.73$ & Terr $\approx 0.81$ & $2.81$ & $1.1\times10^{-2}$ & $2.7$ & $1.6$ & $1.9$ & $-28.23$ & $4.7\times10^{-1}$ & $9.5\times10^{-3}$ & $0.1$ \\
    \hline
    \multirow{3}{*}{500} & $1$ & $1268058741.868$ & Terr $\approx 1$ & $104.3$ & $6.6\times10^{-4}$ & $5.0$ & $3.8$ & $8.3\times10^{-1}$ & $392.732$ & $2.6$ & $1.4$ & $0.06$ \\
    & $2$ & $1266807848.740$ & Terr $\approx 1$ & $9.1$ & $3.4\times10^{-2}$ & $2051.2$ & $1.4$ & $1.6$ & $314.86$ & $1.98$ & $1.36$ & $0.03$ \\
    & $3$ & $1268747361.180$ & Terr $\approx 1$ & $63.9$ & $4.3\times10^{-1}$ & $12.1$ & $1.3$ & $2.0$ & $-163.180$ & $3.1$ & $0.73$ & $0.03$ \\
    \hline
\end{tabular}}
 
\caption{Properties of the most significant foreground associations with a) ANTARES and b) ORCA6. $\Delta t_\mathrm{max}$: maximum time delay. $M_{\rm tot, bank}$: total mass corresponding to the best-fit waveform in PyCBC template bank. $Q_g$: GW Bayes Factor. $E_{\rm rec}$: event reconstructed energy in ANTARES or ORCA6. $Q_{\nu}$: Neutrino Bayes Factor. $\Delta t$: time delay between the neutrino and the GW trigger. $I_{\Omega}$: sky overlap value. $\Lambda$: association rank value. IFAR: Joint Inverse False Alarm Rate [yr]. "Source class." is the GW source classification that can be either BBH, NSBH, BNS, or Terrestrial (noted Terr) here.}
\label{table:ANTARES-ORCA6-foreground30s-assoc}
\end{table*}
\end{landscape}

\subsection{Results}
\label{sec:foreground}

\begin{figure}
    \centering
    \includegraphics[width=0.6\textwidth]{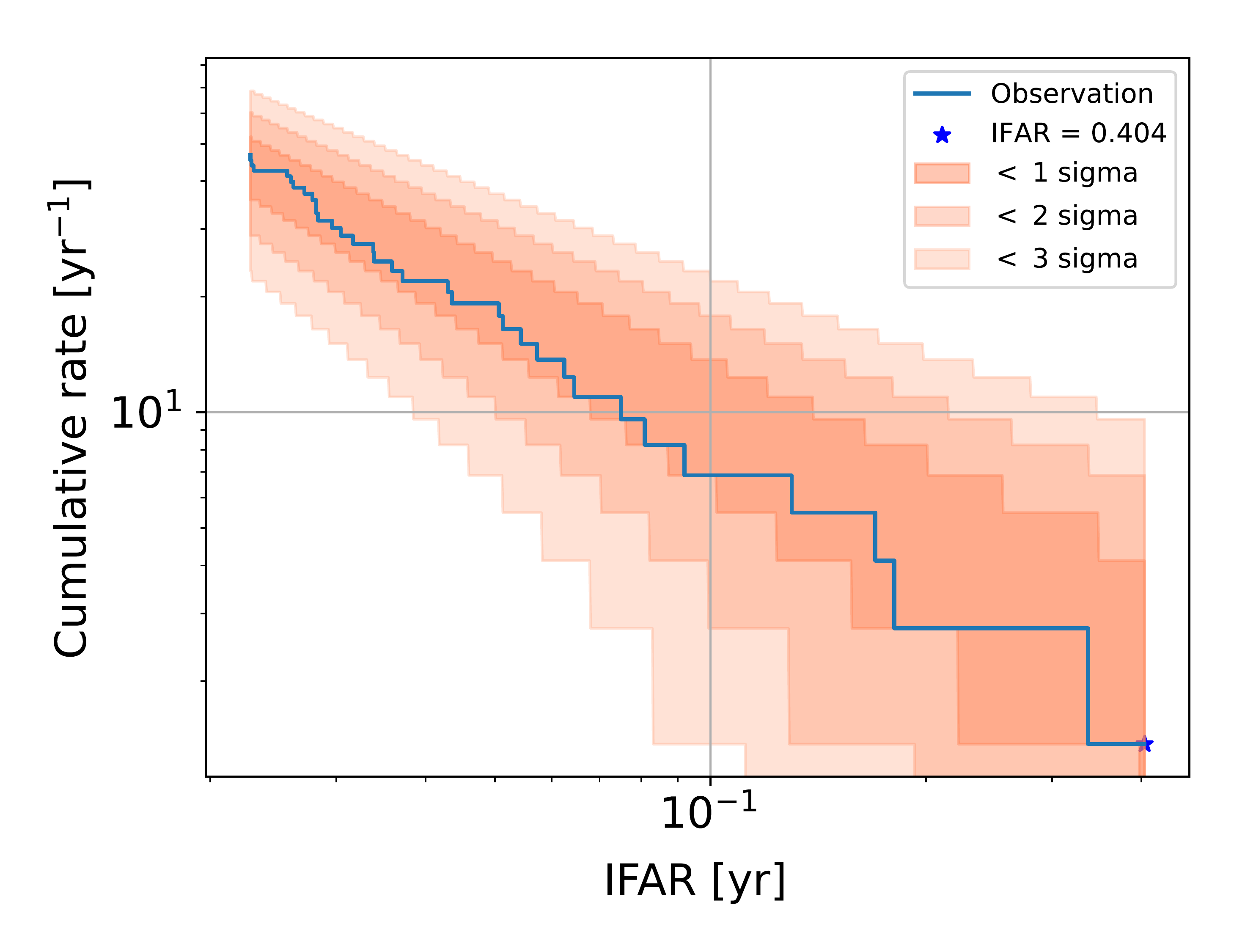}
    \includegraphics[width=0.6\textwidth]{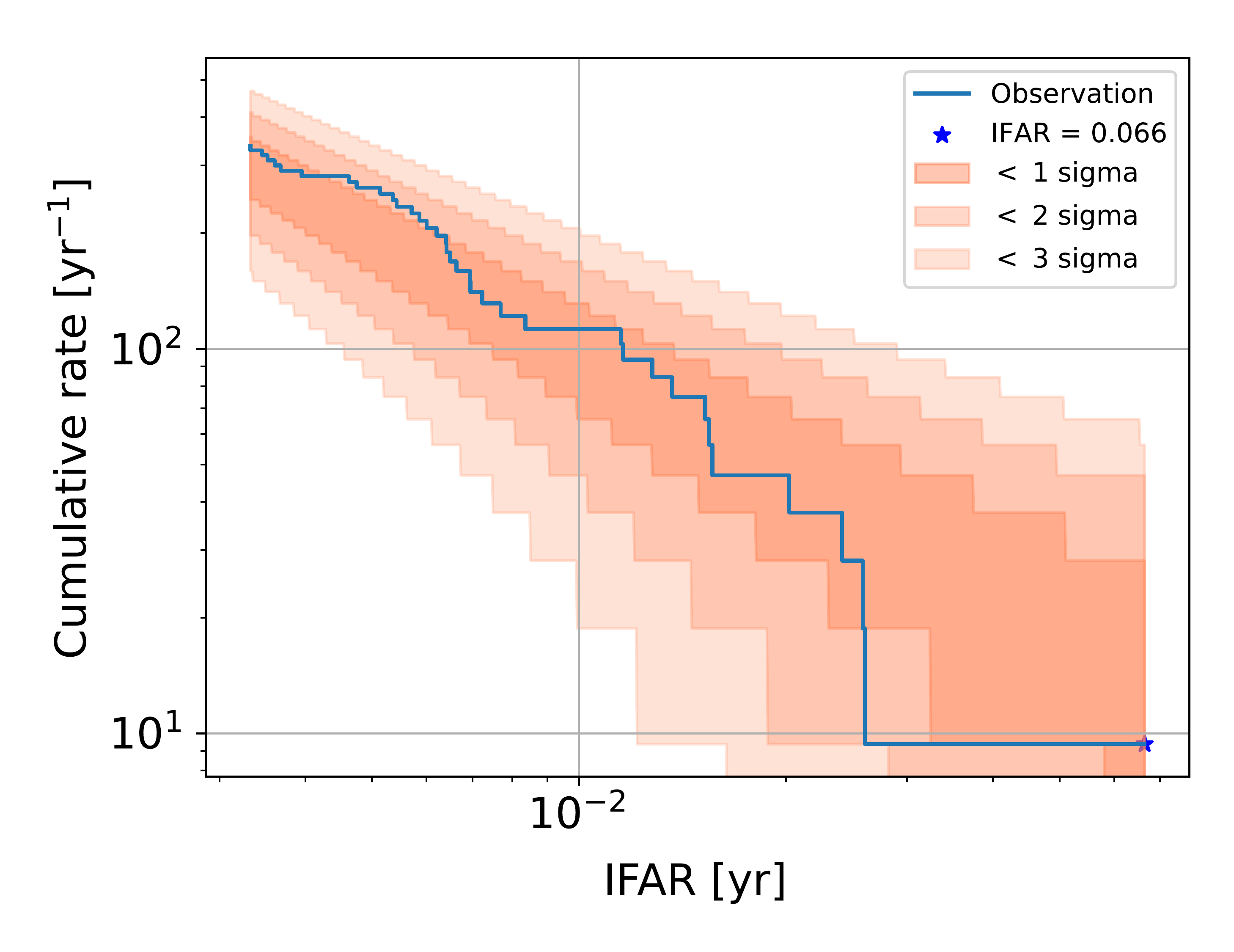}
    \caption{Cumulative rate as a function of the IFAR for the foreground associations obtained by the $\Delta t=\qty{500}{\second}$ analyses with ANTARES (top) and ORCA6 (bottom).}
    \label{fig:ANTARES/ORCA-discovery}
\end{figure}

We apply the method described in the previous sections to GW and neutrino data taken during the O3 period. For the $\Delta t_\mathrm{max} = \qty{500}{\second}$ analysis, the numbers of associations observed by combining LIGO-Virgo with ANTARES and ORCA6 observations are $34$ and $36$, respectively. The cumulative rates of these associations are shown as a function of the IFAR in Figure~\ref{fig:ANTARES/ORCA-discovery}. For both the ANTARES and ORCA6 analyses, the cumulative rate slightly exceeds $1\sigma$ of the background expectation, computed assuming a Poissonian distribution. The IFARs of the most significant foreground associations are $\qty{0.4}{\year}$ and $\qty{0.07}{\year}$ for the ANTARES and ORCA6 analyses, respectively. The skymaps and spectrograms of the corresponding events are shown in Appendix~\ref{foreground-500s-display-appendix}.

\begin{figure*}
    \includegraphics[width=\textwidth]{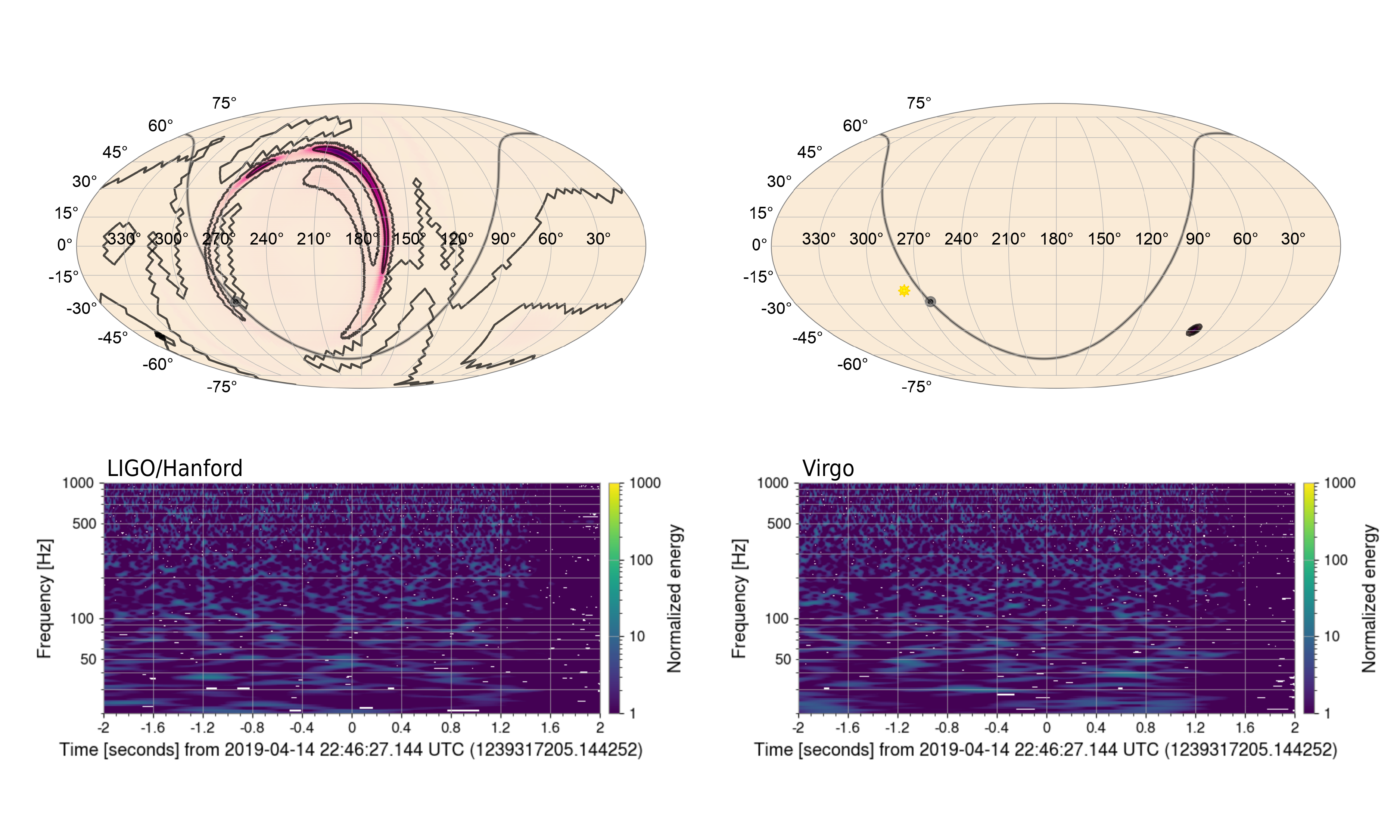}
    \caption{Display of the most significant foreground association with the $\Delta t_\mathrm{max} = \qty{30}{\second}$ analysis in ANTARES. Top: Sky localization in equatorial coordinates of the GW (left) and the neutrino (right) candidates. The three contour levels represent the $3\sigma$, 2$\sigma$, and 1$\sigma$ credible regions. The gray solid line and dot represent the galactic plane and center, respectively. The Sun is represented by the yellow star. The purple scale represents the probability for the event to originate at a given sky localization. Bottom: LIGO/Hanford (left) and Virgo (right) GW spectrogram.}
    \label{fig:1st_foreground_ANTARES_30s}
\end{figure*}

\begin{figure*}
    \includegraphics[width=\textwidth]{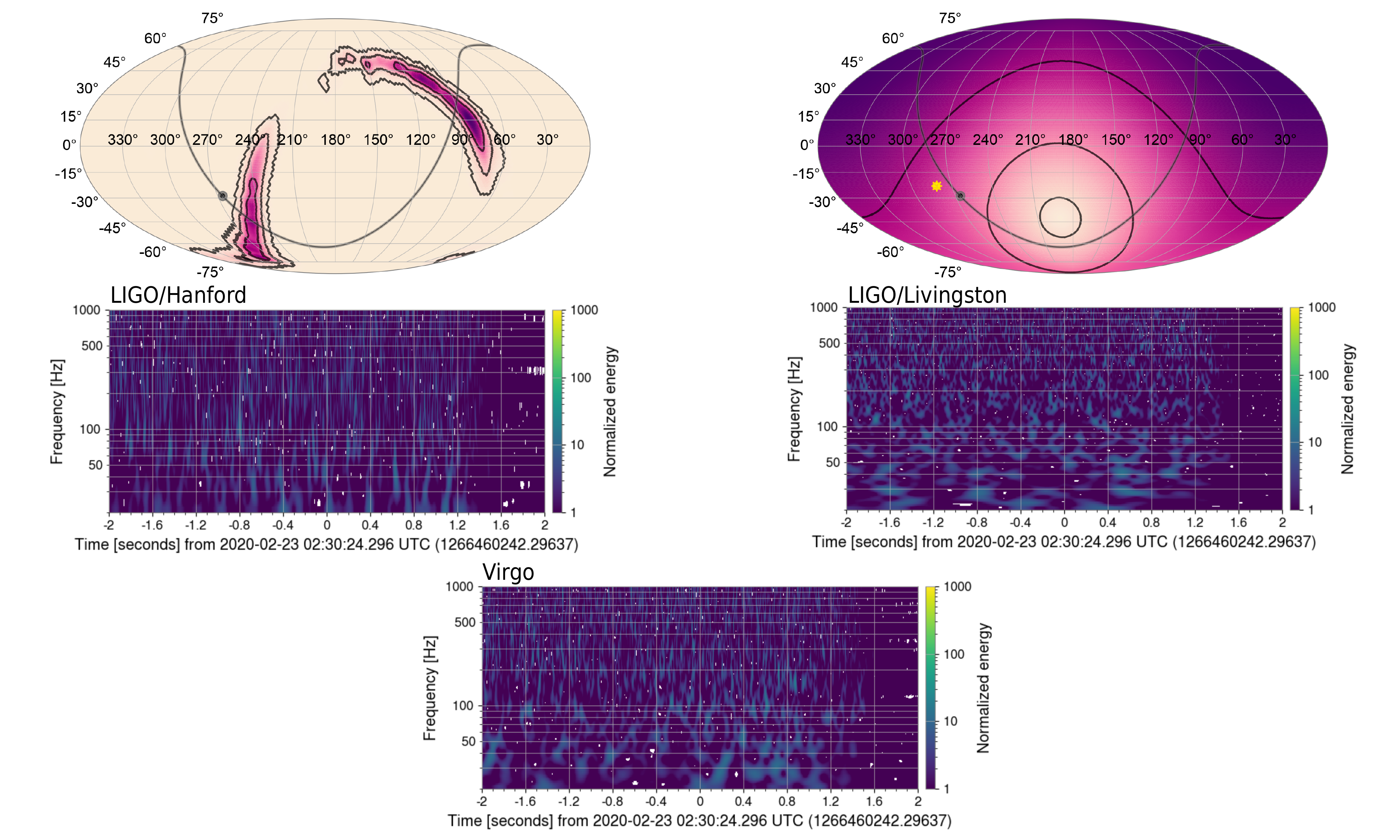}
    \caption{Display of the most significant foreground association with the $\Delta t_\mathrm{max} = \qty{30}{\second}$ analysis in ORCA6. First row: Sky localization of the GW (left) and the neutrino (right) candidates. The three contour levels represent the $3\sigma$, 2$\sigma$, and 1$\sigma$ credible regions. The gray solid line and dot represent the galactic plane and center, respectively. The Sun is represented by the yellow star. The purple scale represents the probability for the event to originate at a given sky localization. Second row: LIGO/Hanford (left) and LIGO/Livingston (right) GW spectrograms. Third row: Virgo spectrogram.}
    \label{fig:1st_foreground_ORCA6_30s}
\end{figure*}

For the $\Delta t_\mathrm{max} = \qty{30}{\second}$ analysis, $3$ and $2$ associations have been identified by the ANTARES and ORCA6 searches, respectively, and we therefore do not consider cumulative rate distributions. The most significant associations for ANTARES and ORCA6 have IFARs of, respectively, $\qty{0.8}{\year}$ and $\qty{0.2}{\year}$. The skymaps and spectrograms for these associations are shown in Figures~\ref{fig:1st_foreground_ANTARES_30s} (ANTARES) and \ref{fig:1st_foreground_ORCA6_30s} (ORCA6).

The properties of the most significant associations for both the $\Delta t_\mathrm{max} = \qty{30}{\second}$ and $\Delta t_\mathrm{max} = \qty{500}{\second}$ analyses are shown in Table~\ref{table:ANTARES-ORCA6-foreground30s-assoc}. Overall, these associations are composed of signal-like GW candidates (small $Q_g$), although a terrestrial origin is favored, in coincidence with a noise-like ($Q_\nu\gtrsim 1$), or sometimes faintly signal-like, neutrino trigger. The sky overlap $I_\Omega$ is generally small and, for ORCA6, reaches large values ($I_\Omega \approx 2-3$) only for poorly-localized neutrino events (see e.g., the skymap in Figure~\ref{fig:1st_foreground_ORCA6_500s}). Due to these limitations, none of these joint GW-HEN observations is significant enough (IFAR $\gtrsim \qty{1}{\year}$) to be claimed as an interesting association. 

\section{Population constraints}
\label{population}
Given that we did not find any plausible joint detection, neither with the triggered search nor the joint analysis, we use both searches to place constraints on the rate of common detectable GW-HEN sources. These population constraints complement the constraints derived by ANTARES and KM3NeT in~\cite{Albert_2023,KM3NeT:2023cdr} for specific GW sources. This section describes the methodology where we consider statistical limits under nominal detector and astrophysical assumptions.

\subsection{Constraints from the triggered search}
\label{subsec:upperlimits_triggered}

The exclusion distances evaluated in Section~\ref{sec:triggered} for different waveforms can be used to estimate upper limits on the rate densities of the corresponding GW-HEN sources. For each waveform model, an exclusion volume can be defined by considering the minimal exclusion distance over all HEN triggers, $D_{90, \mathrm{min}}$. Given this exclusion distance, if the four HEN triggers considered are indeed due to astrophysical neutrinos, the total isotropic energy emitted in neutrinos $E_{\rm iso,\nu}$ corresponding to one expected neutrino is of the order of $10^{54-57}$\,erg, assuming the same spectral index as in Equation~\ref{eq:nuflux}; these energies are far larger than typical predictions for binary mergers. Then, assuming that no more than $1$ GW-HEN event was identified by X-pipeline, and using Poisson statistics, the Neyman 90\% upper limit on the number of expected GW-HEN events from sources located within this volume is $N_\mathrm{90} = 3.9$. The upper limit on the source rate density $R_\mathrm{UL}$ can then be approximated by using

\begin{align}
N_\mathrm{90} = R_\mathrm{UL} \times\frac{ V \times T_\mathrm{obs}}{f_\mathrm{b}},
\label{eq:R_ul}
\end{align}
where $V = 4 \pi D_{90,\mathrm{min}}^3/3$ is the exclusion volume, $T_\mathrm{obs}=228.6$~days is the interferometers' effective observation time---requiring that at least two interferometers are functioning---and $f_\mathrm{b}$ is the beaming factor of the neutrino emission. Here, the beaming factor is set to $f_\mathrm{b} = 10$, corresponding to an opening angle typical for GRBs, as in the previous GW-HEN study~\cite{ANTARES:2018bmu}. 

The rate upper limits for the models studied in Section~\ref{sec:triggered} are shown in Table~\ref{table:waveforms_exclusion_distances}. Additionally, the values of $R_\mathrm{UL}$ for the BNS and NSBH models are shown in Figure~\ref{fig:RUL-ANTARES-ORCA6} and compared to the BNS merger rate. For these models, the amount of energy emitted in GWs, $E_\mathrm{GW}$, has been estimated to lie in the $\qtyrange{0.1}{0.2}{\solarmass}c^2$ range for the NS and BH masses considered. The isotropic energy emitted in neutrinos for the objects considered in the triggered search, equivalent to one expected neutrino in ANTARES, is $E_{\rm iso,\nu} \sim 2.5\times 10^{54}$~erg for BNS and $E_{\rm iso,\nu} \sim 1.2\times 10^{55}$~erg for NSBH mergers. For all models, the upper limits on the source rate density are well above the expected values for the corresponding populations.

\subsection{Constraints from the joint search}

Here, we follow~\cite{ANTARES:2018bmu} rewriting eq.~\ref{eq:R_ul} as:
\begin{equation}
    R_\mathrm{UL} = \frac{3.9 f_\mathrm{b}}{T_\mathrm{obs}} \left[ \int_0 ^\infty 4 \pi r^2 p_\mathrm{det}dr \right]^{-1},
\end{equation}
where $T_\mathrm{obs}= 266.47$ days, and $p_\mathrm{det}$ is defined as:
\begin{equation}
\begin{split}
    p_\mathrm{det} = & f_\mathrm{GW,\nu A}~.~ p_\mathrm{det, \nu A} + f_\mathrm{GW,\nu  K}~.~ p_\mathrm{det, \nu K}  \\ &  - f_\mathrm{GW,\nu A}~.~f_\mathrm{GW,\nu K}~.~p_\mathrm{det, \nu A}~.~p_\mathrm{det, \nu K}
    \end{split}
\end{equation} 
where the index GW designates the LVK dataset and $\nu X$ denotes either the ANTARES ($A$) or ORCA6 ($K$) datasets, and $p_\mathrm{det}$ represents the detection efficiency, estimated from simulations. The parameter $f_\mathrm{GW,\nu X}$ is defined as the fraction of GW-HEN associations with a root-sum-squared GW strain amplitude above a certain threshold (here chosen to correspond to an IFAR of 10 yr), and $p_\mathrm{det, \nu X}$ (with $X$ being either ANTARES or ORCA6) is written as follows:
\begin{equation}
    p_\mathrm{det,\nu X} = 1 - \mathrm{Poisson}(0, n_{\nu X} ),
\end{equation}
where $n_{\nu X}$ is the total number of emitted neutrinos computed from simulations of detector $X$ with a generation spectrum at Earth following $dn_\nu/dE_\nu = \phi_0 E_\nu^{-2.5}$ with $\phi_0$ such that the total isotropic energy emitted in neutrinos, $E_{\rm iso,\nu}$ is $6\times4 \pi r^2 \times \int_{E_{thr}}^\infty E\frac{dn}{dE} \, dE$ with $r$ the distance to the source.

\subsection{Results}

\begin{figure}
    \centering
    \includegraphics[width=0.99\textwidth]{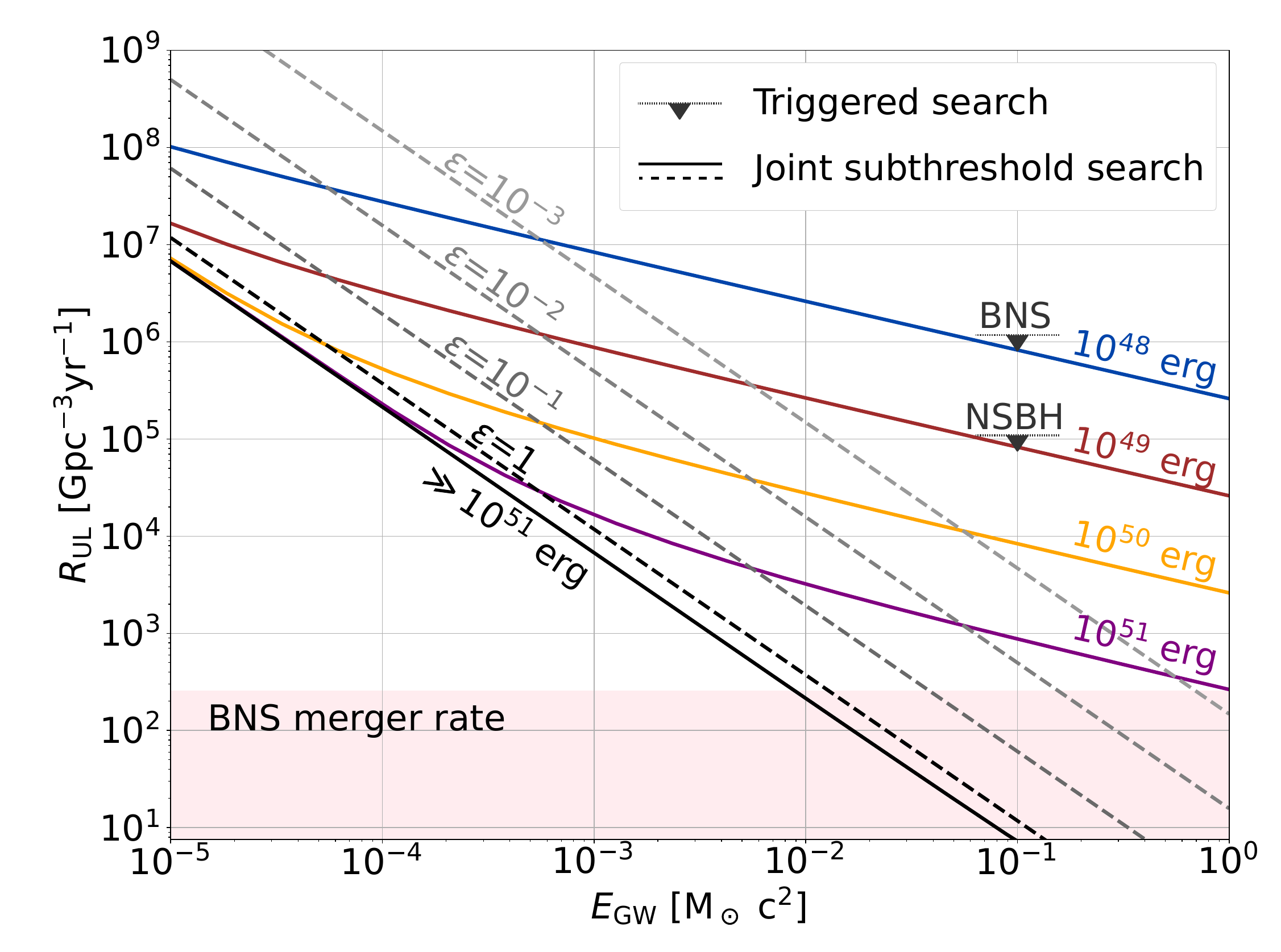}
    \caption{Rate density upper limits for the GW-HEN sources as a function of $E_\mathrm{GW}$, for different values of $E_{\rm iso,\nu}$ (color solid curves) and $\epsilon=E_{\rm iso,\nu}/E_\mathrm{GW}$ (from black to gray dashed curves). The results are obtained using both searches with ANTARES and ORCA6 data (triangles for triggered and lines for subthreshold searches).  For comparison, we also show the local BNS merger rate in pink~\cite{LIGOScientific:2025pvj}. The isotropic energy emitted in neutrinos for the objects considered in the triggered search, equivalent to one expected neutrino in ANTARES, is $E_{\rm iso,\nu} \sim 2.5\times 10^{54}$~erg for BNS and $E_{\rm iso,\nu} \sim 1.2\times 10^{55}$~erg for NSBH mergers.}
    \label{fig:RUL-ANTARES-ORCA6}
\end{figure}

Final results are presented in Figure~\ref{fig:RUL-ANTARES-ORCA6}, both for the triggered search and the subthreshold joint search. For the latter, we present the rate density upper limit of GW-HEN sources for different values of the total energy emitted in GWs and neutrinos. We also show the rate density upper limit curves for different values of $\epsilon = E_{\rm iso,\nu} / E_\mathrm{GW}$. These limits can be compared to those obtained by the triggered search for binary inspirals. In particular, over the $E_\mathrm{GW}$ range corresponding to the BNS and NSBH models described in Section~\ref{sec:injections}, the upper limits on $R_\mathrm{UL}$ from the subthreshold search match those of the triggered search for neutrinos energies of $10^{48}$ and $10^{49}$~ergs for BNS and NSBH mergers, respectively, six orders of magnitude below the minimal neutrino energies considered for the triggered search. We can also compare these limits to those obtained in previous studies, such as \cite{ANTARES:2018bmu}. 
For a typical BNS with a GW emitted energy of $E_\mathrm{GW} = \qty{e-1}{\solarmass}c^2$, we find a joint source rate upper limit of \qty{8.8e2}{\per\cubic\giga\parsec\per\year} considering an isotropic energy emitted in neutrinos of $E_{\rm iso,\nu} = \qty{e51}{\erg}$. Although this upper limit remains above the predicted BNS rate range~\cite{LIGOScientific:2025pvj}, it improves over the previous limit of  \qty{6e3}{\per\cubic\giga\parsec\per\year} \cite{ANTARES:2018bmu} by a factor of $7$.

\section{Conclusion}
\label{sec:conclusion}

Two multi-messenger analyses were performed in order to search for GW events in coincidence with HEN triggers using the data taken by the LIGO-Virgo GW detectors and by the ANTARES and KM3NeT/ORCA6 neutrino experiments during the O3 period. The triggered analysis described in Section~\ref{sec:triggered}, aimed at the joint detection of GW-HEN emission using \texttt{X-pipeline}, exploiting the information of four selected HEN triggers. Our results led to no joint detections, with all the GW candidate \textit{p}-values consistent with the null joint detection hypothesis. As no coincidence was found, we computed for all HEN the $90\%$ lower limits on the distance to the GW-HEN sources for different models, which are reported in Table~\ref{table:waveforms_exclusion_distances}. Since the parameters of the simulated GW waveforms are taken to be similar to the ones used in previous searches for GW+GRB coincidences~\cite{o3afrb,Abbott2017ApJ84189A}, the exclusion distances found for the SGC and ADI models ($< \qty{200}{\mega\parsec}$) are in line with the values reported in these searches. The triggered search performed in this paper will be extended to future analyses with GW-HEN data. 

Besides the HEN-triggered search, we adapted the joint multi-messenger analysis described in \cite{Pillas_2023}, designed to find subthreshold GW candidates in LIGO/Virgo data in coincidence with data from the KM3NeT and ANTARES neutrino detectors. We ranked the potential associations using the ranking statistic described in \citep{Ranking}, based on the time and spatial overlap and the significance of the GW and the neutrino candidates. To check the validity of the analysis, we also applied it to simulated associations composed of GW simulated signals injected in O3 data and temporally and spatially coincident with simulated astrophysical neutrinos, assuming an $E^{-2.5}$ spectrum. About 90\% of the simulated associations with ANTARES and ORCA6 are recovered with an IFAR above 1 yr, and the most significant simulated associations have an IFAR $>\qty{4153}{\year}$ (ANTARES) and $>\qty{610}{\year}$ (KM3NeT/ORCA6). Differences mostly arise from the duty cycle difference between the two detectors. Simulated associations with IFAR lower than 1 yr are mostly explained by transient GW detector noise overlapping with the injections, or by a single-detector GW injection with a large sky localization uncertainty. Applying the search to O3 data from LIGO/Virgo and to neutrino data, considering ANTARES and ORCA6 separately, yielded no significant association with an IFAR higher than \qty{0.8}{\year}. This non-detection result is consistent with conclusions from previous searches applied to O3 data~\citep{Abbasi_2023, Albert_2023, KM3NeT:2023cdr}. Finally, we used our simulations and our non-detection to obtain upper limits on the rate of common GW-HEN sources. Applying the method from \cite{ANTARES:2018bmu} to the ANTARES and ORCA6 data, we constrained this rate to be below \qty{4e3}{\per\cubic\giga\parsec\per\year} for sources with $E_\mathrm{GW} = \qty{e-2}{\solarmass}c^2$ and $E_{\rm iso,\nu} = \qty{e51}{\erg}$, which is $10$ times lower than previous results from \cite{ANTARES:2018bmu}. These rate upper limits are consistent with the ones from the triggered search results. 

Both searches are planned to be applied to the data collected during the fourth GW observing run (O4). During O4, ANTARES was no longer running, but both KM3NeT/ORCA and KM3NeT/ARCA had considerably expanded compared to O3~\citep{KM3NeT:2024paj}. With $21$ to $30$ lines deployed throughout O4, KM3NeT/ARCA, in particular, had reached a higher instrumented volume than ANTARES for a similar energy range. Given the improvement of the GW and neutrino detector sensitivities, we thus expect to considerably improve the constraints presented in this article, and possibly detect subthreshold GW-HEN associations within LIGO/Virgo and KM3NeT data.

\begin{acknowledgments}
The authors acknowledge the financial support of:
%INFRADEV
KM3NeT-INFRADEV2 project, funded by the European Union Horizon Europe Research and Innovation Programme under grant agreement No 101079679;
%Belgium
Funds for Scientific Research (FRS-FNRS), Francqui foundation, BAEF foundation.
%Czeck
Czech Science Foundation (GAČR 24-12702S);
%France
Agence Nationale de la Recherche (contract ANR-15-CE31-0020), Centre National de la Recherche Scientifique (CNRS), Commission Europ\'eenne (FEDER fund and Marie Curie Program), LabEx UnivEarthS (ANR-10-LABX-0023 and ANR-18-IDEX-0001), Paris \^Ile-de-France Region, Normandy Region (Alpha, Blue-waves and Neptune), France,
%For the CPER
The Provence-Alpes-Côte d'Azur Delegation for Research and Innovation (DRARI), the Provence-Alpes-Côte d'Azur region, the Bouches-du-Rhône Departmental Council, the Metropolis of Aix-Marseille Provence and the City of Marseille through the CPER 2021-2027 NEUMED project,
%For IN2P3
The CNRS Institut National de Physique Nucléaire et de Physique des Particules (IN2P3),
%for ACME
The ACME project funded by the European Union’s Horizon Europe Research and innovation programme under Grant Agreement No 101131928;
%Georgia
Shota Rustaveli National Science Foundation of Georgia (SRNSFG, FR-22-13708), Georgia;
%Germany (Max Planck Inst.)
ERC MuSES project No 101142396); ERC starting grant MessMapp, under contract No. 949555.
%Greece
The General Secretariat of Research and Innovation (GSRI), Greece;
%Italy
Istituto Nazionale di Fisica Nucleare (INFN) and Ministero dell’Universit{\`a} e della Ricerca (MUR). KM3NeT4RR MUR Project National Recovery and Resilience Plan (NRRP), Mission 4 Component 2 Investment 3.1, Funded by the European Union – NextGenerationEU,CUP I57G21000040001, Concession Decree MUR No. n. Prot. 123 del 21/06/2022;
%Morocco
Ministry of Higher Education, Scientific Research and Innovation, Morocco, and the Arab Fund for Economic and Social Development, Kuwait;
%The Netherlands
Nederlandse organisatie voor Wetenschappelijk Onderzoek (NWO), the Netherlands;
%Poland
The grant “AstroCeNT: Particle Astrophysics Science and Technology Centre”, carried out within the International Research Agendas programme of the Foundation for Polish Science financed by the European Union under the European Regional Development Fund; The program: “Excellence initiative-research university” for the AGH University in Krakow; The ARTIQ project: UMO-2021/01/2/ST6/00004 and ARTIQ/0004/2021;
%Romania
Ministry of Education and Scientific Research, Romania
%Slovak Republic
Slovak Research and Development Agency under Contract No. APVV-22-0413; Ministry of Education, Research, Development and Youth of the Slovak Republic;
%Spain
MICIU for PID2024-156285NB-C41, -C42- C43, funded by MICIU/AEI/10.13039/501100011033 and by FEDER, EU, and for CNS2023-144099; Generalitat Valenciana for CIDEGENT/2020/049, CIDEGENT/2021/23, CIDEIG/2023/20, CIPROM/2023/51 and INNVA1/2024/110 (IVACE+i), and Fundaci\'{o}n Bancaria La Caixa (ID 100010434), for LCF/BQ/PI25/12100025, Spain;
%UAE
Khalifa University internal grants (ESIG-2023-008, RIG-2023-070 and RIG-2024-047), United Arab Emirates;
%UK
The European Union's Horizon 2020 Research and Innovation Programme (ChETEC-INFRA - Project no. 101008324).
% disclaimer
Views and opinions expressed are those of the author(s) only and do not necessarily reflect those of the European Union or the European Research Council. Neither the European Union nor the granting authority can be held responsible for them. The authors are grateful for computational resources provided by the LIGO Laboratory and supported by National Science Foundation Grants PHY-0757058 and PHY-0823459.
\end{acknowledgments}

\bibliographystyle{JHEP}
\bibliography{reference}

\appendix

\section{ANTARES and ORCA6 data distribution}
\label{dataset}
\begin{figure}[hbtp]
    \centering
    \includegraphics[width=\linewidth]{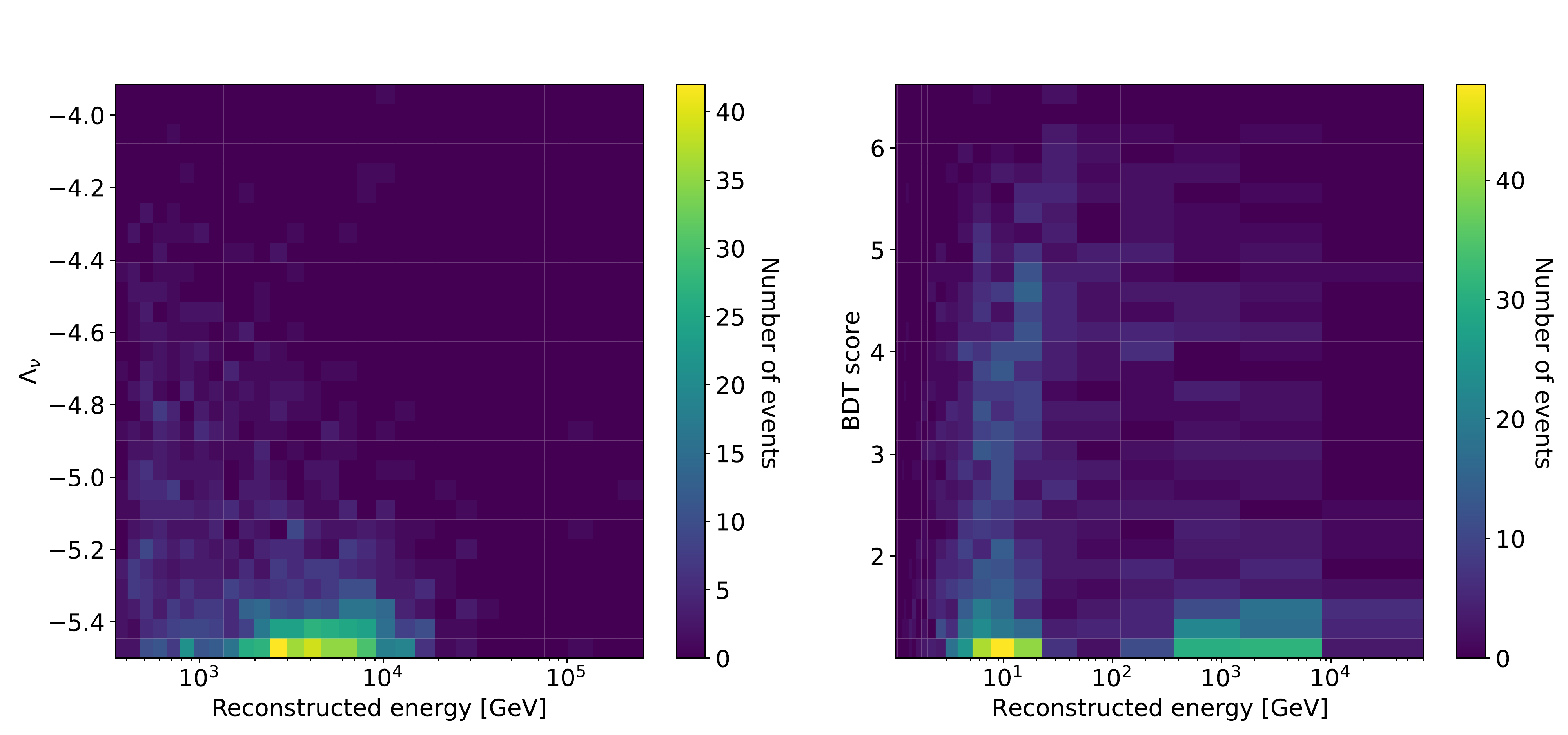}
    \caption{Left: Distribution of ANTARES reconstructed events during O3 as a function of the reconstructed energy and the quality factor $\Lambda_\nu$.
    Right: Distribution of ORCA6 reconstructed events during O3 as a function of the reconstructed energy and the BDT score.}
    \label{fig:ANTARES-ORCA6-data-distrib}
\end{figure}

The Figure~\ref{fig:ANTARES-ORCA6-data-distrib} (left) shows the distribution of ANTARES reconstructed events during O3 as a function of the reconstructed energy and the quality factor $\Lambda_\nu$ (for reconstructed events with $\Lambda_\nu~>~-5.5$). The corresponding distribution of ORCA6 data during O3 as a function of the reconstructed energy and the BDT score is shown in Figure~\ref{fig:ANTARES-ORCA6-data-distrib} (right). Atmospheric muons are concentrated at low BDT scores and $\Lambda_\nu$ values. The expected fraction of astrophysical neutrinos increases with the reconstructed energy and the reconstruction quality.

\section{Limitations of the simulated association significance}
\label{sec:limitations}

\begin{figure*}
    \centering
    \includegraphics[width=0.7\textwidth]{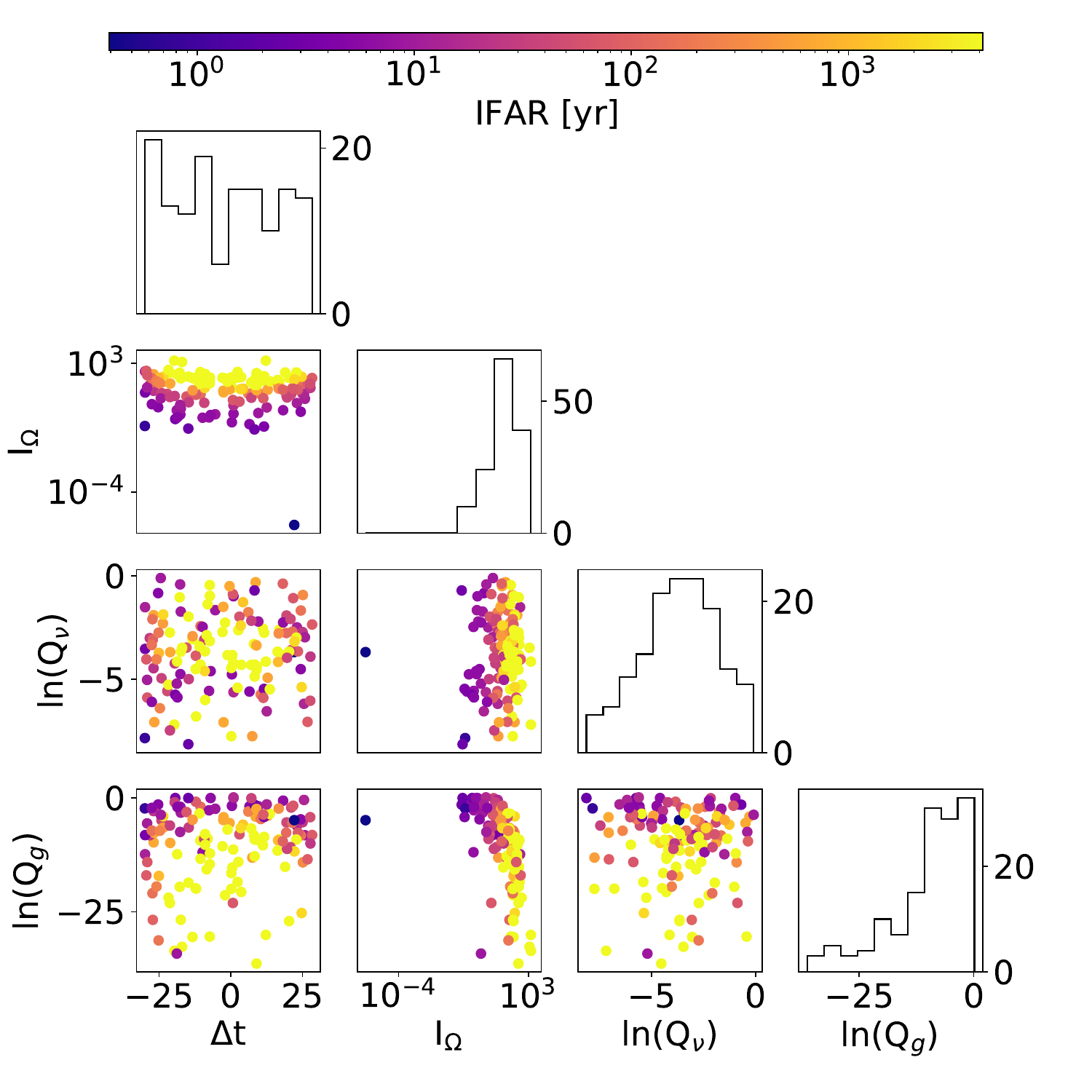}
    \includegraphics[width=0.7\textwidth]{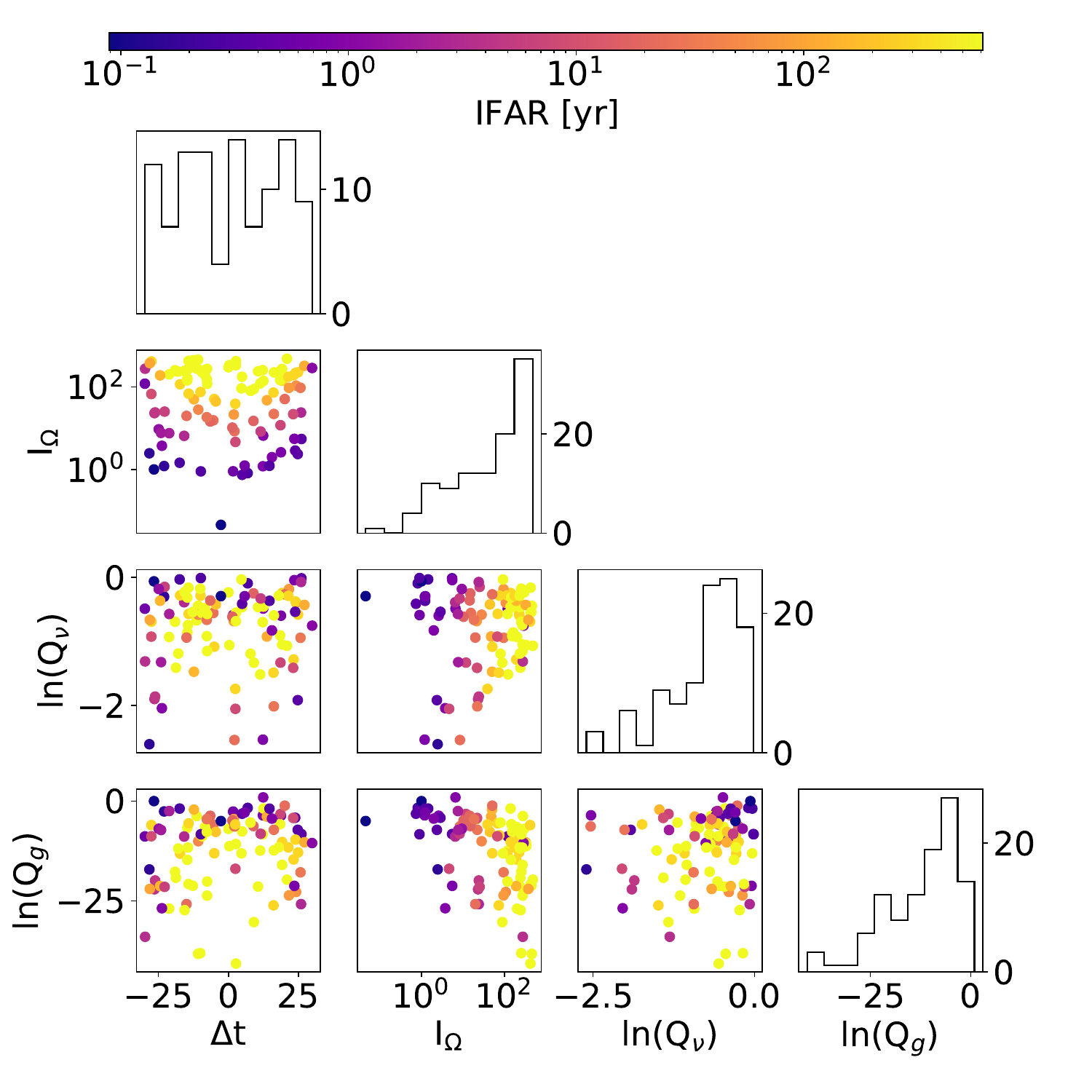}
    \vspace*{-0.4cm}
    \caption{Distribution of the simulated associations as a function of the various Bayes factors for the $\Delta t_\mathrm{max} = \qty{30}{\second}$ search with ANTARES (top) and ORCA6 (bottom) data.
    $Q_g$: GW Bayes Factor. $Q_{\nu}$: Neutrino Bayes Factor. $\Delta t$: time delay between the neutrino and the GW trigger. $I_{\Omega}$: sky overlap value.}
    \label{fig:corner-injections}
\end{figure*}

\begin{figure*}
    \centering
    \includegraphics[width=0.7\textwidth]{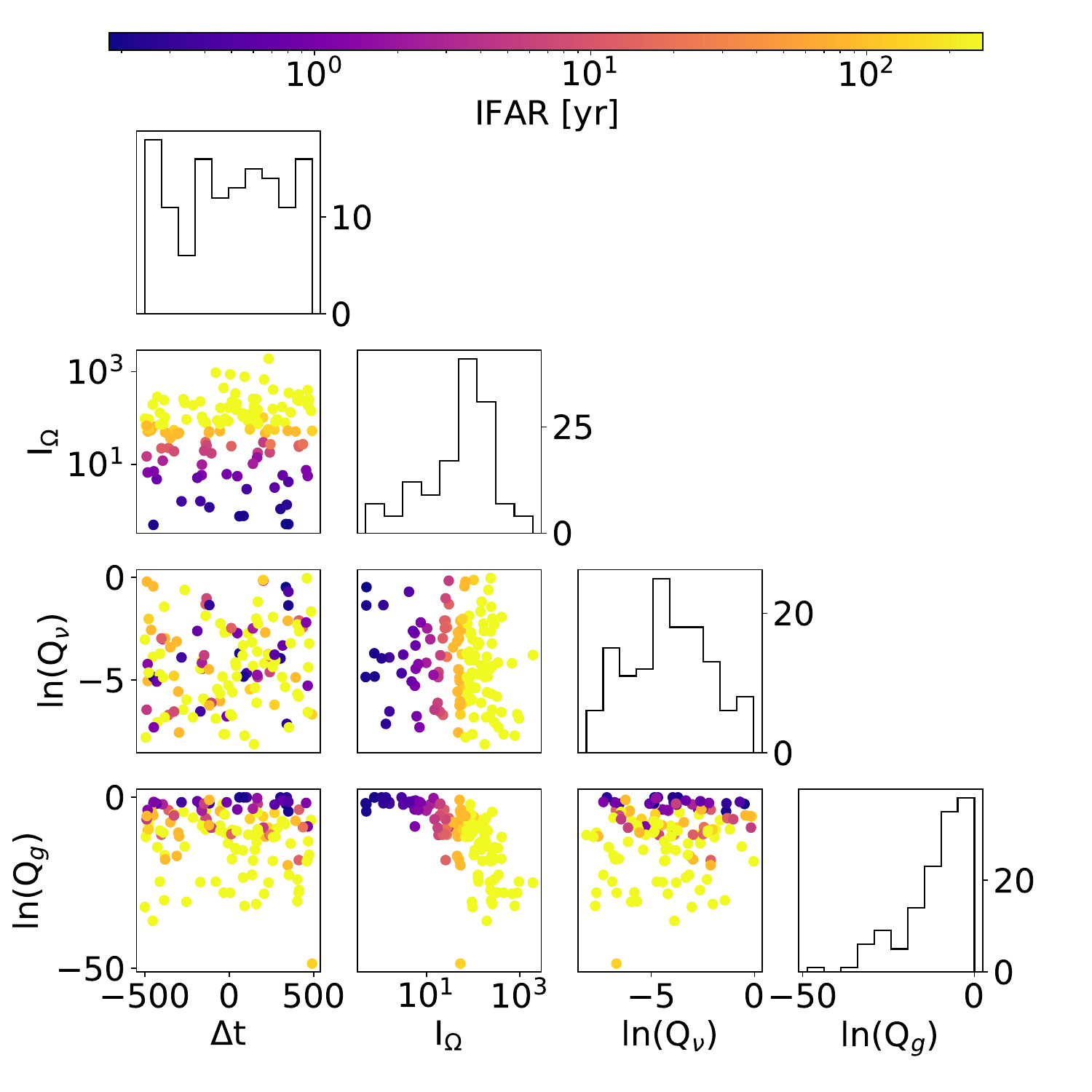}
    \includegraphics[width=0.7\textwidth]{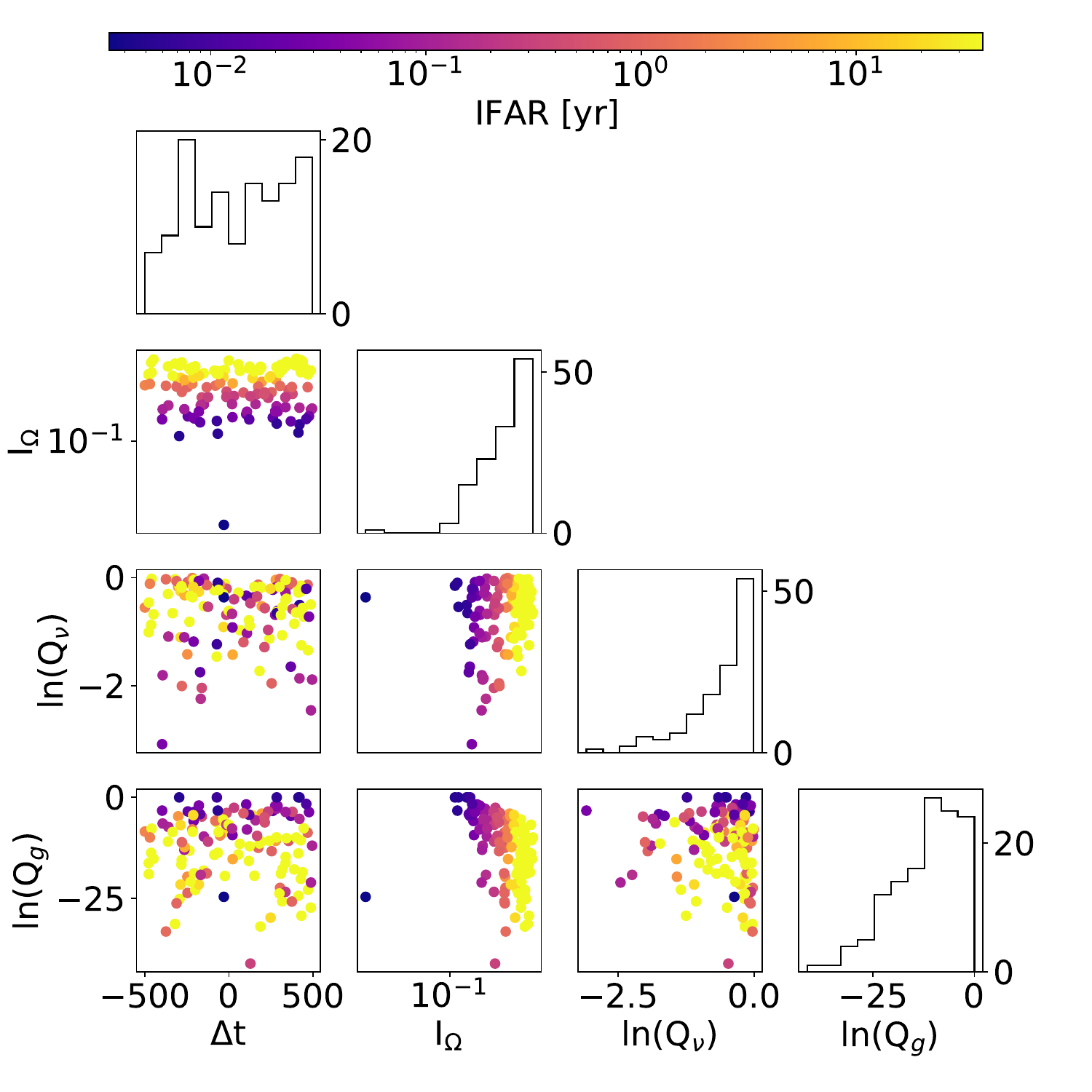}
    \caption{Distribution of the simulated associations as a function of the various Bayes factors for the $\Delta t_\mathrm{max} = \qty{500}{\second}$ search, for ANTARES (top) and ORCA6 (bottom).
    $Q_g$: GW Bayes Factor. $Q_{\nu}$: Neutrino Bayes Factor. $\Delta t$: time delay between the neutrino and the GW trigger. $I_{\Omega}$: sky overlap value.}
    \label{fig:corner-injections-500s}
\end{figure*}

We investigate the correlations between the IFAR and the Bayes factors computed in the ranking statistic for all the injections discussed in Section~\ref{sec:injections}. These correlations are shown through corner plots in Figure~\ref{fig:corner-injections} for the $\Delta t_\mathrm{max} = \qty{30}{\second}$ analysis in ANTARES (top) and ORCA6 (bottom). The corresponding figure for the $\Delta t_\mathrm{max} = \qty{500}{\second}$ analysis is shown in Figure~\ref{fig:corner-injections-500s}. Each point represents one simulated association and is colored by its IFAR value. The IFAR strongly depends on the sky overlap term, and the simulated signals with a small $I_\Omega$ are not well recovered. Other correlations confirm that the method is efficient: small time delay associations have, in general, higher IFAR than those with large time delays, and small $Q_g$ and $Q_\nu$ (signal-like) associations give higher IFAR than coincidences with large GW and neutrino trigger significance Bayes factors. 

For the $\Delta t_\mathrm{max} = \qty{30}{\second}$ search, over the 30 associations with an IFAR below 1 yr, half of them are composed of at least one candidate (either GW or neutrino) with a sky localization uncertainty area above \qty{e4}{\square\deg} ($\sim$ all-sky), which must significantly reduce the sky overlap Bayes factor.

In the following paragraph, we focus on the $\Delta t_\mathrm{max} = \qty{30}{\second}$ search, and we investigate the correlations between the IFAR and the sky and time overlap Bayes factors.
Smaller $I_\Omega$ and larger time delays between the GW and neutrino candidate are associated with lower significances (see first column -- fourth row panel of Figure~\ref{fig:corner-injections}).

\textit{Associations with small $I_{\Delta t}$}: As the time delay has been randomly chosen between \qty{-30}{\second} and \qty{+30}{\second}, it is not surprising that some associations have smaller IFAR because of small $I_{\Delta t}$, as, in this analysis, the closer in time the two candidates are, the higher the probability that they come from a common source.

\begin{figure*}
    \centering
    \includegraphics[width=\linewidth]{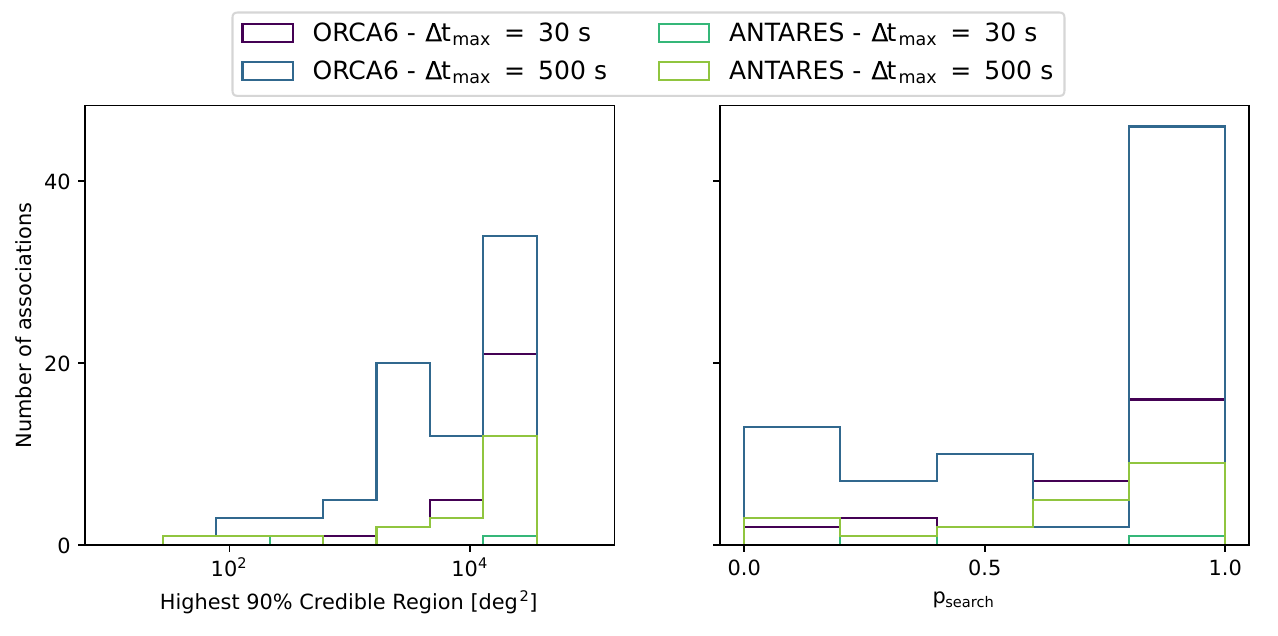}
    \caption{Left: Distribution of the largest localization uncertainty within each GW–HEN association with a joint IFAR below 1 yr. Right: Distribution of $p_{\rm search}$ estimated for the GW localization for associations with a joint IFAR below 1 yr.}
    \label{fig:investigations-GW-skyloc}
\end{figure*}

\textit{Associations with small $I_{\Omega}$}: These associations can have two different origins: i) the sky localization uncertainty of either the GW injection or the MC astrophysical neutrino is extremely large; ii) the reconstructed 90\% contour sky localization of these GW injections does not contain the injected (true) sky localization. This second scenario can happen when loud glitches in the data overlap with GW signals, or when very loud GW injections, close and heavy, are mistaken for glitches. Indeed, the angular separation between the true and reconstructed sky localization of the MC neutrino has been restricted to be below \qty{5}{\degree}, but there is no such requirement for the GW injection. To investigate the first case, the distribution of the largest sky localization uncertainty between the GW and the neutrino simulated signals is shown in Figure~\ref{fig:investigations-GW-skyloc} (left panel), for associations with a joint IFAR below \qty{1}{\year}. The maximum localization uncertainty can reach values larger than $\qty{e4}{\square\deg}$, either because only one GW interferometer was working at the time of the injection, or because of a large uncertainty on the neutrino direction. The latter can indeed be observed for ORCA6 events in Figure~\ref{fig:angular-sep-MC}.
For the second case, we define $p_\mathrm{search}$ as the cumulative probability within the smallest highest-posterior-density region that contains the true location. As shown in the right panel of Figure~\ref{fig:investigations-GW-skyloc}, more than half of the injections with an IFAR $< \qty{1}{\year}$ have a large $p_\mathrm{search}$ value, indicating that the injected direction is in a low-posterior-density region.

\section{Most significant foreground association with the $\Delta t_{\mathrm{max}} = \qty{500}{\second}$ analysis}
\label{foreground-500s-display-appendix}

\begin{figure*}
    \includegraphics[width=\textwidth]{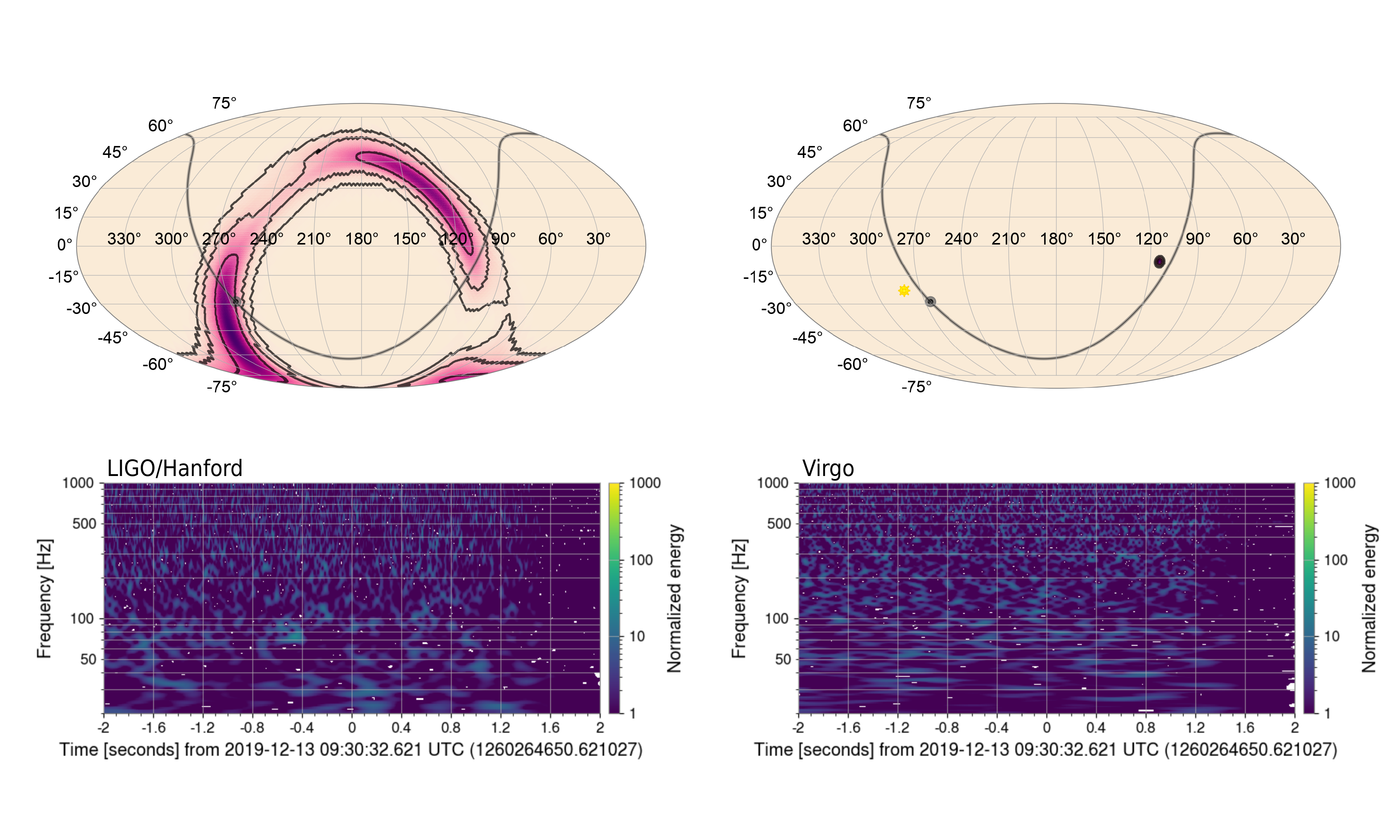}
    \caption{Display of the most significant foreground association with the $\Delta t_\mathrm{max} = \qty{500}{\second}$ analysis in ANTARES. Top: Sky localization of the GW (left) and the neutrino (right) candidates. The three contour levels represent the $3\sigma$, 2$\sigma$, and 1$\sigma$ credible regions. The gray solid line and dot represent the galactic plane and center, respectively. The Sun is represented by the yellow star. The purple scale represents the probability for the event to originate at a given sky localization. Bottom: LIGO/Hanford (left) and Virgo (right) GW spectrograms.}
    \label{fig:1st_foreground_ANTARES_500s}
\end{figure*}

\begin{figure*}
    \includegraphics[width=\textwidth]{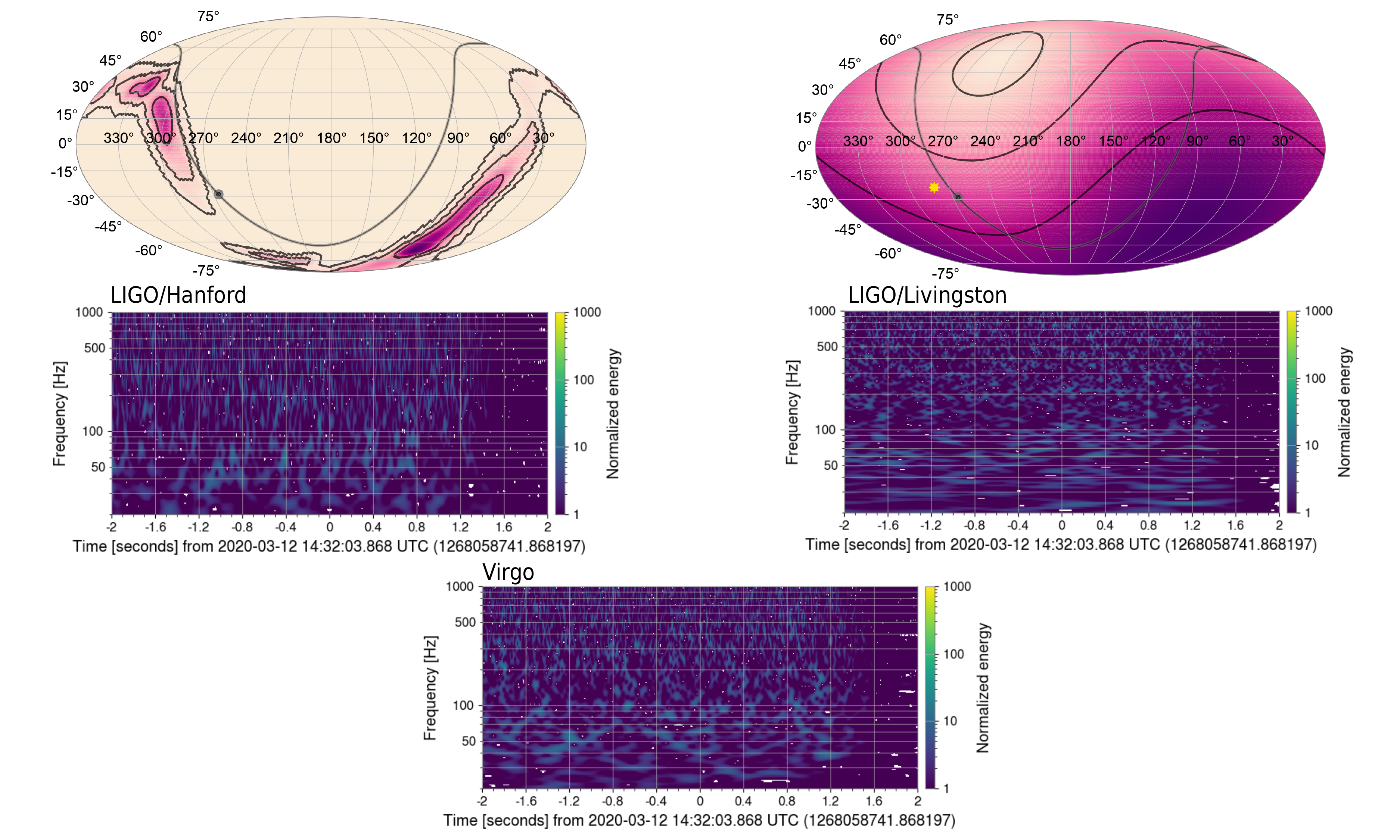}
    \caption{Display of the most significant foreground association with the $\Delta t_\mathrm{max} = \qty{500}{\second}$ analysis in ORCA6. First row: Sky localization of the GW (left) and the neutrino (right) candidates. The three contour levels represent the $3\sigma$, 2$\sigma$, and 1$\sigma$ credible regions. The gray solid line and dot represent the galactic plane and center, respectively. The Sun is represented by the yellow star. The purple scale represents the probability for the event to originate at a given sky localization. Second row: LIGO/Hanford (left) and LIGO/Livingston (right) GW spectrograms. Third row: Virgo spectrogram.}
    \label{fig:1st_foreground_ORCA6_500s}
\end{figure*}

Figures~\ref{fig:1st_foreground_ANTARES_500s} and \ref{fig:1st_foreground_ORCA6_500s} show the display of the most significant foreground association with the $\Delta t_\mathrm{max} = \qty{500}{\second}$ analysis for ANTARES and ORCA6, respectively (similarly to Figures~\ref{fig:1st_foreground_ANTARES_30s} and \ref{fig:1st_foreground_ORCA6_30s} for the $\Delta t_\mathrm{max} = \qty{30}{\second}$ analysis).

\end{document}